\documentclass[journal]{IEEEtran}
\usepackage{hyperref}
\usepackage{balance}
\usepackage{amsmath}
\usepackage{array}
\usepackage{graphicx}
\usepackage{amssymb}
\usepackage{pifont}
\usepackage{subfigure}
\usepackage{booktabs}
\usepackage{multirow}
\usepackage{color}
\usepackage{caption}
\usepackage{bm}
\usepackage{balance}
\newcommand{\cmark}{\ding{51}}
\newcommand{\xmark}{\ding{55}}

\ifCLASSINFOpdf
\else
\fi
\begin{document}

\title{DSRGAN: Detail Prior-Assisted Perceptual \\ Single Image Super-Resolution via \\Generative Adversarial Networks}
\author{Ziyang Liu,
        Zhengguo Li,
        Xingming Wu,
        Zhong Liu, and
        Weihai Chen*
\thanks{*Corresponding author: Weihai Chen, whchen@buaa.edu.cn.}
\thanks{Z. Liu, X. Wu, Z. Liu, and W. Chen are with School of Automation Science and Electrical Engineering, Beihang University, 100191, Beijing, China.}
\thanks{Z. Li is with SRO Department, Institute for Infocomm Research, Singapore, 138632, Singapore.}}

\markboth{}%
{Shell \MakeLowercase{\textit{et al.}}: Bare Demo of IEEEtran.cls for IEEE Journals}
\maketitle
\begin{abstract}
The generative adversarial network (GAN) is successfully applied to study the perceptual single image super-resolution (SISR). However, the GAN often tends to generate images with high frequency details being inconsistent with the real ones.
Inspired by conventional detail enhancement algorithms, we propose a novel prior knowledge, the detail prior, to assist the GAN in alleviating this problem and restoring more realistic details. The proposed method, named DSRGAN, includes a well designed detail extraction algorithm to capture the most important high frequency information from images.
Then, two discriminators are utilized for supervision on image-domain and detail-domain restorations, respectively.
The DSRGAN merges the restored detail into the final output via a detail enhancement manner. The special design of DSRGAN takes advantages from both the model-based conventional algorithm and the data-driven deep learning network.
Experimental results demonstrate that the DSRGAN outperforms the state-of-the-art SISR methods on perceptual metrics and achieves comparable results in terms of fidelity metrics simultaneously.
Following the DSRGAN, it is feasible to incorporate other conventional image processing algorithms into a deep learning network to form a model-based deep SISR.
\end{abstract}

\IEEEpeerreviewmaketitle

\section{Introduction}
\IEEEPARstart{G}{enerating} a high-resolution (HR) image from one single low-resolution (LR) image, which is called single image super-resolution (SISR), is a classic and key problem in the field of image processing \cite{yang2010image}. The generated image is referred to as the super-resolution (SR) image. Recently most SISR methods are deep learning-based. They can be divided into two categories according to their orientations to different image quality metrics: fidelity metrics such as the peak signal-to-noise ratio (PSNR) and structural similarity index (SSIM), and perceptual metrics such as the perceptual index (PI) \cite{pirm} and learned perceptual image patch similarity (LPIPS) \cite{zhang2018unreasonable}.

The fidelity-oriented SISR methods are commonly formalized as the minimization of the mean squared error (MSE) or mean absolute error (MAE) between the recovered SR image and the real HR image \cite{srcnn}.
Higher PSNR and SSIM have been achieved by these methods with the development of convolutional neural network (CNN) \cite{vdsr,edsr,wdsr}.
However, the minimization of the Euclidean distance between predicted and ground truth pixels will encourage the CNN to average all plausible outputs and produce blurry results, where edges and textures are smoothed.

\begin{figure}[t]
  \centering
  \begin{minipage}[b]{1\linewidth}
  \centering

    \begin{minipage}[b]{0.24\linewidth}
      \centering
      \includegraphics[width=2.1cm,height=1.4cm]{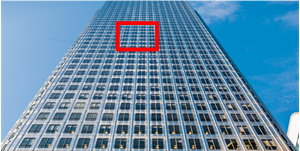}\vspace{-3pt}
      \centerline{\footnotesize{(a) img\_030}}\vspace{2pt}
    \end{minipage}
    \begin{minipage}[b]{0.24\linewidth}
      \centering
      \includegraphics[width=2.1cm,height=1.4cm]{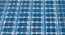}\vspace{-3pt}
      \centerline{\footnotesize{(b) HR}}\vspace{2pt}
    \end{minipage}
        \begin{minipage}[b]{0.24\linewidth}
      \centering
      \includegraphics[width=2.1cm,height=1.4cm]{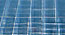}\vspace{-3pt}
      \centerline{\footnotesize{(c) SRGAN}}\vspace{2pt}
    \end{minipage}
        \begin{minipage}[b]{0.24\linewidth}
      \centering
      \includegraphics[width=2.1cm,height=1.4cm]{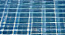}\vspace{-3pt}
      \centerline{\footnotesize{(d) ESRGAN}}\vspace{2pt}
    \end{minipage}

    \end{minipage}

  \begin{minipage}[b]{1\linewidth}
  \centering

    \begin{minipage}[b]{0.24\linewidth}
      \centering
      \includegraphics[width=2.1cm,height=1.4cm]{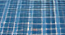}\vspace{-3pt}
      \centerline{\footnotesize{(e) SPSR}}\vspace{0pt}
    \end{minipage}
    \begin{minipage}[b]{0.24\linewidth}
      \centering
      \includegraphics[width=2.1cm,height=1.4cm]{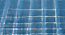}\vspace{-3pt}
      \centerline{\footnotesize{(f) RankSRGAN}}\vspace{0pt}
    \end{minipage}
        \begin{minipage}[b]{0.24\linewidth}
      \centering
      \includegraphics[width=2.1cm,height=1.4cm]{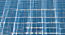}\vspace{-3pt}
      \centerline{\footnotesize{(g) HCFlow}}\vspace{0pt}
    \end{minipage}
        \begin{minipage}[b]{0.24\linewidth}
      \centering
      \includegraphics[width=2.1cm,height=1.4cm]{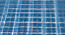}\vspace{-3pt}
      \centerline{\footnotesize{(h) DSRGAN}}\vspace{0pt}
    \end{minipage}

      \end{minipage}

  \vfill
  \caption{SR results of different perception-oriented SISR methods. Our proposed DSRGAN preserves more sharp and geometric-consistent edges than existing methods including the SRGAN \cite{srgan}, ESRGAN \cite{esrgan}, SPSR \cite{ma2020structure}, RankSRGAN \cite{ranksrgan}, and HCFlow \cite{hcflow}.}

 \label{fig:introduction}
\end{figure}

The generative adversarial network (GAN) \cite{gan} has been developed to tackle this problem. The generator and discriminator in GAN are formalized to play a two-player min-max game. By playing the game, the generator network is capable of producing images with rich details, which are visual-pleasing and photo-realistic \cite{srgan}. The GAN has been widely adopted to study the perception-oriented SISR \cite{esrgan, ma2020structure, ranksrgan}. Although these methods perform well on the PI and LPIPS, they often tend to produce distortions in SR images. Their performances on the PSNR and SSIM are unsatisfactory, which implies that the details generated by the generator are different from the real ones. As shown in Figures \ref{fig:introduction}(c) and \ref{fig:introduction}(d), a few edges restored by the SRGAN \cite{srgan} and ESRGAN \cite{esrgan} are twisted. The reason is that the discriminator in GAN may introduce unstable factors to the optimization procedure \cite{ma2020structure}. In addition, the ill-posed nature of the SISR problem makes the generator difficult to restore details consistent with the real ones.

Since restoring the real high frequency details becomes the toughest problem in the SISR, a detail prior-assisted SISR paradigm via the GAN, named DSRGAN, is proposed to alleviate this issue.
The proposal is inspired by a few conventional detail enhancement algorithms \cite{wls, fwls, wang2019detail, kou2017intelligent, wei2016local, farbman2008edge, gif, wgif}, according to which a natural image can be decomposed into a base layer and a detail layer. The base layer is formed by homogeneous image regions with overly smoothed textures. The detail layer reveals the sharp edges of each local region and the rich textures within it. We pay more attention to the detail layer and exploit it to explicitly guide the perception-oriented SISR networks to enhance the SR performance.

Specifically, the DSRGAN includes a novel model-based detail extraction algorithm to extract the detail layer. The algorithm is able to capture the most important high frequency information of image, such as sharp edges and fine textures. It can also intelligently distinguish and exclude unwanted information such as noises, which obstruct the SISR reconstruction. Moreover, the algorithm contains a regularization term to ensure a sparse attribute of the detail layer, such that it can guide the network more efficiently. Once the detail layer is extracted from the LR image, it serves as a prior knowledge for the generator in DSRGAN. The DSRGAN includes two discriminators, one for supervision on image-domain LR-to-HR translation and the other for supervision on detail-domain LR-to-HR translation. Finally, the DSRGAN merges the restored detail into the SR image via a detail enhancement manner \cite{wls, wang2019detail}. The special design of DSRGAN takes advantages from both the model-based conventional algorithm and the data-driven deep learning network.
As such, it is capable of restoring more realistic details, as shown in Figure \ref{fig:introduction}(h). Notably, the DSRGAN is network-agnostic, which can be used for off-the-shelf SISR networks. Furthermore, the detail extraction algorithm is not required to be differentiable. Following the DSRGAN, it is also feasible to incorporate other conventional image processing algorithms into a deep network to form a model-based deep SISR, which will have a great potential research value. In summary, our main contributions are:
\begin{itemize}
\item A novel detail extraction algorithm is proposed to exploit the detail prior for explicitly guiding the perception-oriented SISR reconstruction.
\item A detail prior-assisted SISR paradigm via the GAN, named DSRGAN, is proposed, which takes advantages from both the model-based conventional algorithm and the data-driven deep learning network.
\item The proposed DSRGAN achieves the best on perceptual metrics compared with the most recent state-of-the-art SISR methods, meanwhile brings minor distortions to fidelity metrics.
\end{itemize}

The rest of this paper is organized as below. Relevant works on the SISR are presented in Section II. Details on the proposed DSRGAN are given in Section III. Extensive experimental results are provided in Section IV to verify the performance of DSRGAN. Finally, conclusion remarks are provided in Section V.

\section{Related work}
In this section, existing fidelity-oriented and perception-oriented SISR methods are reviewed. The SISR methods assisted by different kinds of prior are also summarized.
\subsection{Fidelity-oriented SISR methods}
In the field of deep learning-based SISR, Dong et al. firstly proposed a simple three-layer CNN to realize the SISR end-to-end \cite{srcnn}. Subsequently, networks of various architectures were proposed for a higher restoration accuracy \cite{vdsr,fsrcnn,espcn,drrn,edsr,wdsr}. These methods target at achieving high PSNR and SSIM by adopting L1 or L2 regularizations as loss functions. However, the fidelity-oriented SISR methods tend to average plausible outputs and produce blurry results, where edges and textures are smoothed.

\subsection{Perception-oriented SISR methods}
It is important to preserve the sharp edges and fine textures such that the SR image looks more realistic. To achieve this, Johnson et al. proposed a loss function defined on the feature-domain to measure the perceptual differences between SR and HR images \cite{loss}. Ledig et al. proposed a GAN framework, named SRGAN, to perform a deterministic LR-to-HR image translation \cite{srgan}. Then, the feature-domain loss and GAN are widely studied for producing SR images with high visual quality as perceived by human observers.
Sajjadi et al. proposed a texture loss inspired by the feature-domain loss \cite{enhancenet}. Wang et al. introduced a relativistic GAN to better measure the distribution differences between SR and HR images \cite{esrgan}. Zhang et al. proposed a rank-content loss for optimizing network parameters in the direction of better no-reference perceptual metrics \cite{ranksrgan}. Shaham et al. proposed an internal learning scheme to conduct the SISR reconstruction by learning from a single LR image \cite{singan}. Shaham et al. disabled the feature-domain loss for preserving higher PSNR and SSIM. Fuoli et al. proposed a fourier space loss to encourage the network to recover high frequency information \cite{Fuoli}. Lugmayr et al. presented the normalizing flow to predict the diverse photo-realistic HR images instead of using GAN \cite{srflow, hcflow}. 

\subsection{Prior-assisted SISR methods}
Although perception-oriented SISR methods are able to produce SR images of high perceptual quality, they sometimes generate unnatural artifacts and fake high frequency details. Recently, a few methods attempted to exploit extra image prior knowledge to alleviate the ill-posed SR problem and to restore high frequency information closer to the original HR images. The image gradient prior is the most widely used kind of prior. Yang et al. employed an off-the-shelf edge detector to extract the edge gradient, which was then used to guide the deep network to reconstruct SR images with more clear edges \cite{yang2017deep}. Ma et al. proposed a gradient-domain loss based on a differentiable gradient extraction module, to help the network concentrate more on geometric structures and suppress undesired structural distortions \cite{ma2020structure}. Semantic prior can also be a guide for the SR reconstruction. Wang et al. proposed to pre-compute the semantic information of the LR image, and then utilized the condition mechanism to generate textures and details that are more in line with their corresponding semantic categories \cite{sftgan}. Li et al. introduced the face semantic information to achieve better SR reconstruction for face images \cite{li2020learning}.

Different from these methods, we introduce a novel detail prior to guide the deep network to restore SR images with more realistic details, which can be of higher perceptual quality. Compared with the gradient prior, the detail prior preserves not only edge profile information, but also finer texture information. To the best of our knowledge, no perception-oriented SISR methods have explicitly exploited detail information for the SR reconstruction.

\section{The proposed DSRGAN}
\begin{figure*}[t]
\centering
\includegraphics[width=160mm]{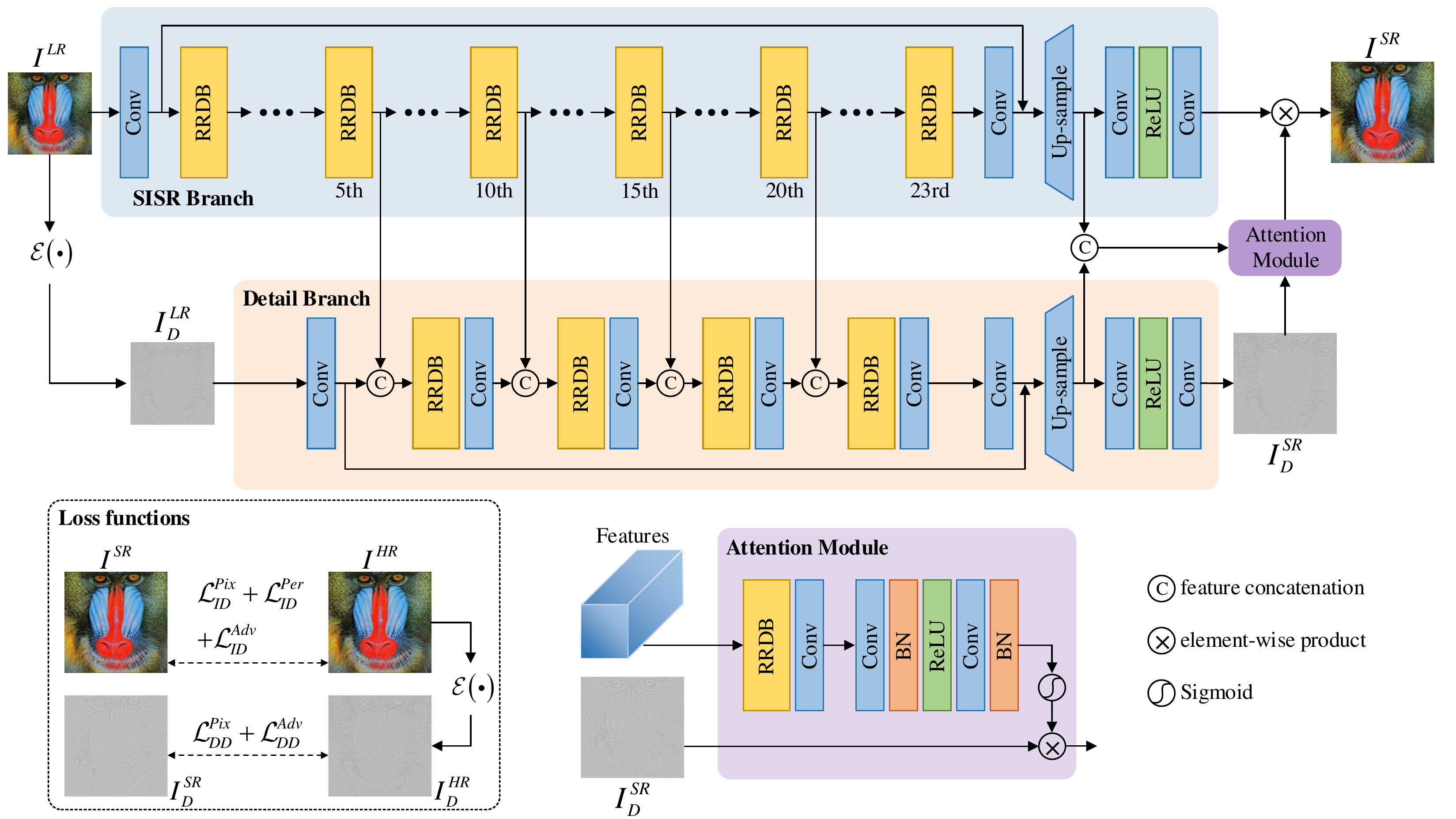}
\caption{The overal framework of the proposed DSRGAN. It includes a novel model-based detail extraction algorithm, a generator, an image-domain discriminator, and a detail-domain discriminator. The generator network is composed of a main SISR branch, an auxiliary detail branch, and an attention module.}
\label{fig:srnet}
\end{figure*}

The overall framework of DSRGAN is shown in Figure \ref{fig:srnet}. It includes a novel detail extraction algorithm, a generator network, and two discriminator networks. In this section, we give details on each part of the DSRGAN and present how it combines the model-based conventional algorithm and the data-driven deep learning network.

\subsection{Detail extraction}
In the field of image processing, an image $I$ is often required to be decomposed into a base layer $I_B$ and a detail layer $I_D$. 
$I_B$ captures the image structure information and is mainly formed by homogeneous regions. While $I_D$ carries the high frequency information and is constituted of residual smaller scale details. We focus on $I_D$ as restoring high frequency details is the most critical part of the SISR reconstruction.
Inspired by our previous work \cite{wls}, a detail extraction algorithm is proposed to acquire $I_D$ for the SISR.
The detail extraction is divided into two stages: 1) generation of a vector field; 2) extraction of fine details from the vector field.

The RGB image $I$ is firstly converted into the YCbCr image and the luminance component $Y$ is extracted. Then, the vector field $(V_h, V_v)$ (including horizontal and vertical dimensions) is constructed by computing gradients in \textit{log} domain as:
\begin{align}
\label{eq:vector1}
V_h(i,j)&=(1+\alpha)\log_2(\frac{Y(i+1,j)+1}{Y(i,j)+1}),\\ \label{eq:vector2}
V_v(i,j)&= (1+\alpha)\log_2(\frac{Y(i,j+1)+1}{Y(i,j)+1}),
\end{align}
where $(i,j)$ represents the pixel coordinate. $\alpha(\geq 0)$ is a hyper parameter and it amplifies the details such that they have more chances to be extracted. Notably, a large $\alpha$ also amplifies the noises.

The detail layer $I_D$ is extracted from $(V_h, V_v)$ by solving the following quadratic optimization problem \cite{wls}:
\begin{equation}
\label{eq:quadoptimize}
\min_{I_D}\{\|I_D\|^2+\lambda[(\frac{V_h-\frac{\partial I_D}{\partial h}^2}{\psi(V_h)})^2+(\frac{V_v-\frac{\partial I_D}{\partial v}^2}{\psi(V_v)})^2]\},
\end{equation}
where the function $\psi(z)$ is defined as:
\begin{equation}
\label{eq:psi}
\psi(z)=\sqrt{|z|^{\gamma}+\epsilon}.
\end{equation}
The objective (\ref{eq:quadoptimize}) includes a regularization term and a fidelity term. The regularization term $\|I_D\|^2$ encourages the detail layer to be sparse. The sparsity can help the network concentrate more on useful information. The fidelity term preserves the details by measuring the difference between $\nabla I_D$ and $V$. $\lambda$ is a hyper parameter that achieves a tradeoff between two energy terms. The function $\psi$ is to prevent the noises of $V$ from appearing in $I_D$.
$\gamma$ determines the sensitivity of the objective to $V$. $\epsilon$ prevents division by zero and a larger value excludes more noises from $I_D$. Notably, small scale details can also be excluded if $\epsilon$ is too large.

The optimal solution to the objective (\ref{eq:quadoptimize}) can be obtained by solving the following linear equation:
\begin{align}
\nonumber
(E+\lambda (D_h' \mathcal{A}(V_h) & D_h + D_v' \mathcal{A}(V_v) D_v)) I_D\\
& = \lambda (D_h' \mathcal{A}(V_h)V_h + D_v' \mathcal{A}(V_v) V_v),
\label{eq:linearquad}
\end{align}
where $D_h'$ represents the transpose of $D_h$. The function $\mathcal{A}(z)$ is defined as:
\begin{equation}
\label{eq:diag}
\mathcal{A}(z) = \rm{diag}(\frac{1}{\psi^2(z)}).
\end{equation}
$E$ is the identity matrix. $D_h$ and $D_v$ are discrete differentiation operators along the horizontal and vertical dimensions, respectively.

Solving the matrix $(E+\lambda (D_h' \mathcal{A}(V_h) D_h + D_v' \mathcal{A}(V_v) D_v))^{-1}$ is complicated. Different from \cite{wls}, we apply a 1-Dimensional edge-preserving smoothing technique in \cite{fwls} to approximate the matrix via several iterations:
\begin{align}
\nonumber
&(E+\lambda (D_h' \mathcal{A}(V_h) D_h + D_v' \mathcal{A}(V_v) D_v))^{-1} \\
&\approx \prod_{t=1}^{T}(E+\lambda_t (D_h' \mathcal{A}(V_h) D_h))^{-1}(E+ \lambda_t (D_v' \mathcal{A}(V_v) D_v))^{-1},
\label{eq:approximate}
\end{align}
where $T$ represents the number of iterations and $\lambda_t$ is obtained as follows:
\begin{align}
\lambda_t = \frac{3}{2}\frac{4^{T-t}}{4^T-1} \lambda.
\end{align}
In other words, the objective (\ref{eq:quadoptimize}) is decomposed into two 1-Dimensional optimization problems along the horizontal and vertical dimensions, i.e., $I_D^h$ and $I_D^v$, respectively:
\begin{equation}
\label{xdirectionoptimization}
\min_{I_D^h}\sum_{i=0}^{W-1}[(I_D^{h}(i))^2+\lambda_t(\frac{V_h(i)-\frac{\partial I_D^h(i)}{\partial i}}{\psi (V_h(i))})^2],
\end{equation}
\begin{equation}
\label{ydirectionoptimization}
\min_{I_D^v}\sum_{j=0}^{H-1}[(I_D^{v}(j))^2+\lambda_t(\frac{V_v(j)-\frac{\partial I_D^v(j)}{\partial j}}{\psi (V_v(j))})^2],
\end{equation}
where $H$ and $W$ correspond to the height and width of image, respectively. Clearly, solving the matrices $(E+\lambda_t (D_h' \mathcal{A}(V_h) D_h))^{-1}$ and $(E+ \lambda_t (D_v' \mathcal{A}(V_v) D_v))^{-1}$ for the objectives (\ref{xdirectionoptimization}) and (\ref{ydirectionoptimization}) becomes much easier than solving the objective (\ref{eq:quadoptimize}). The matric have an exact optimal solution that can be obtained using the Gaussian elimination algorithm via a complexity of $O(H)$ or $O(W)$. The overall computational complexity of the proposed detail extraction algorithm is $O(T \times H \times W)$, which is greatly reduced compared with \cite{wls}.

Notably, the detail extraction algorithm is different from \cite{farbman2008edge, gif, wgif}, whereas $I_B$ is filtered from $I$ by edge-preserving smoothing algorithms and $I_D$ is obtained by subtracting $I_B$ from $I$. As a result, $I_D$ tend to hold negative values and possible noises that filtered out by smoothing. While in the proposed detail extraction algorithm, a quadratic objective is directly defined on $I$ in \textit{log} domain. After solving the objective, an exponential operation is required for the solution to obtain the final $I_D$ \cite{wls}. Thus, $I_D$ is strictly positive. We define a function $\mathcal{E}$ to denote the entire detail extraction algorithm in this paper, i.e., $I_D=\mathcal{E}(I)$.

\subsection{The generator network of DSRGAN}
The DSRGAN learns the high frequency details in an explicit way with the help of $\mathcal{E}$.
The generator of DSRGAN includes three parts: a main SISR branch, an auxiliary detail branch, and an attention module.

The SISR branch is composed of 23 residual-in-residual dense blocks (RRDBs) \cite{esrgan}. The nearest neighbor interpolation is performed at the end of the branch to resize features into the same size of the HR image $I^{HR}$. Finally, the features are mapped to the image domain via the convolution-ReLU-convolution.

The detail branch takes the detail layer of $I^{LR}$ as input, which is obtained by $I_D^{LR}=\mathcal{E}({I^{LR}})$, and restores its HR counterpart $I_D^{SR}$. The ultimate goal of detail branch is to learn high frequency details explicitly and to feedback the details to the main SISR branch. Following \cite{ma2020structure},
the detail branch is composed of only 4 RRDBs considering the increment of network parameters.
However, a lightweight detail branch is struggling to restore the seriously lost details. Therefore, we incorporate the features from the SISR branch to enhance the representation ability of the detail branch. Specifically, the features from the 5th, 10th, 15th, and 20th RRDBs in the SISR branch are concatenated with the features from the 1st, 2nd, 3rd, and 4th RRDBs in the detail branch, as shown in Figure \ref{fig:srnet}. Each RRDB in the detail branch is followed by a convolution layer to reduce the feature channels. Finally, the upsampled features are mapped to the detail domain via the convolution-ReLU-convolution.

Once the $I_D^{SR}$ is generated,
it is multiplied back to the image domain to generate the final SR image $I^{SR}$. The multiplication is inspired by the conventional detail enhancement algorithm \cite{wls}. 
It is also shown in \cite{wang2019detail} that such detail enhancement can improve both the subjective and the objective qualities of image.

Furthermore, we apply an attention on the $I_D^{SR}$ to prevent artifacts caused by inappropriate detail enhancement. The attention module takes features from both the SISR and detail branches and computes the attention map though a few convolutions, activations, and normalizations. Owing to this design, the restored detail layer can enhance the SR image in a more adaptive way.

We define a function $\mathcal{G}$ to denote the generator of DSRGAN and we get:
\begin{align}
I^{SR}, I_D^{SR} = \mathcal{G}(I^{LR}, I_D^{LR} \ | \ \theta_{\mathcal{G}}), \label{eq:generator}
\end{align}
where $\theta_{\mathcal{G}}$ corresponds to the network parameters.

\subsection{Loss functions}
The loss functions can be divided into image-domain losses and detail-domain losses, respectively. The DSRGAN includes an image-domain discriminator $\mathcal{D}_{ID}$ and a detail-domain discriminator $\mathcal{D}_{DD}$, for supervision on restoring $I^{SR}$ and $I_D^{SR}$, respectively.
\subsubsection{Image-domain loss functions}
Commonly used image-domain loss functions are constituted of a pixel loss, a perceptual loss, and an adversarial loss \cite{srgan, esrgan, ma2020structure}. The pixel loss measures the pixel-wise differences between $I^{SR}$ and ground truth $I^{HR}$:
\begin{align}
\mathcal{L}_{ID}^{Pix} = \mathbb{E}_{I^{SR}}{|| I^{SR} - I^{HR} ||}_1. \label{eq:pix-ID}
\end{align}

The perceptual loss measures the differences between the deep-level features of SR and HR images, which are extracted from a pre-trained VGG model \cite{perceptual}:
\begin{align}
\mathcal{L}_{ID}^{Per} = \mathbb{E}_{I^{SR}}{|| \mathcal{V}(I^{SR}) - \mathcal{V}(I^{HR}) ||}_1, \label{eq:per-ID}
\end{align}
where $\mathcal{V}$ represents the VGG model.

The adversarial loss $\mathcal{L}_{ID}^{Adv}$ includes an energy function for the $\mathcal{G}$ and $\mathcal{D}_{ID}$. They are optimized by playing a two-player game:
\begin{align}
\nonumber
\mathcal{L}_{ID}^{Adv}(\theta_{\mathcal{D}_{ID}}) = & -\mathbb{E}_{I^{SR}}[{\rm{log}}(1-\mathcal{D}_{ID}(I^{SR}))] \\ \label{eq:dis-ID}
&-\mathbb{E}_{I^{HR}}[{\rm{log}}(\mathcal{D}_{ID}(I^{HR}))],
\end{align}
\begin{align}
\mathcal{L}_{ID}^{Adv}(\theta_{\mathcal{G}}) = -\mathbb{E}_{I^{SR}}[{\rm{log}}(\mathcal{D}_{ID}(I^{SR}))], \label{eq:gen-ID}
\end{align}
where $\mathcal{D}_{ID}$ is implemented by a encoder network similar to \cite{srgan, esrgan, ma2020structure}. We also employ a relativistic average GAN proposed in \cite{esrgan} to boost the performance of DSRGAN.

\subsubsection{Detail-domain loss functions}
In typical deep learning-based SISR methods, details are implicitly learned by residual learning, where the input and output of network are skipped connected \cite{vdsr, edsr, wdsr}. However, the deep learning network works in a black-box way, which does not guarantee that the residuals are of high frequency. Different from these works, the DSRGAN learns high frequency information in a more interpretable way. To achieve this, the detail-domain loss functions are proposed for supervision of the detail layer. The loss functions constituted of a pixel loss and an adversarial loss, respectively.

Similar to $\mathcal{L}_{ID}^{Pix}$, the pixel loss defined on the detail domain measures the pixel-wise differences between $I_D^{SR}$ and ground truth $I_D^{HR}$:
\begin{align}
\mathcal{L}_{DD}^{Pix} = \mathbb{E}_{I_D^{SR}}{|| I_D^{SR} - I_D^{HR} ||}_1, \label{eq:pix-DD}
\end{align}
where $I_D^{HR} = \mathcal{E}(I^{HR})$.

A simple L1 regularization can lead the $I_D^{SR}$ to be smooth and it is too weak to capture high frequency differences.
We further employ a powerful GAN to solve this problem. Similar to $\mathcal{L}_{ID}^{Adv}$, a detail-domain discriminator $\mathcal{D}_{DD}$ is introduced and $\mathcal{G}$ and $\mathcal{D}_{DD}$ are optimized by playing another two-player game:
\begin{align}
\nonumber
\mathcal{L}_{DD}^{Adv}(\theta_{\mathcal{D}_{DD}}) = & -\mathbb{E}_{I_D^{SR}}[{\rm{log}}(1-\mathcal{D}_{DD}(I_D^{SR}))] \\ \label{eq:dis-DD}
&-\mathbb{E}_{I_D^{HR}}[{\rm{log}}(\mathcal{D}_{DD}(I_D^{HR}))],
\end{align}
\begin{align}
\mathcal{L}_{DD}^{Adv}(\theta_{\mathcal{G}}) = -\mathbb{E}_{I_D^{SR}}[{\rm{log}}(\mathcal{D}_{DD}(I_D^{SR}))], \label{eq:gen-DD}
\end{align}
where $\mathcal{D}_{DD}$ shares the same network architecture with $\mathcal{D}_{ID}$. In practice, a relativistic average GAN is utilized. Notably, the extracted $I_D^{LR}$ and $I_D^{HR}$ are sparse, which is very friendly to the optimization procedure of $\mathcal{G}$ and $\mathcal{D}_{DD}$.

The overall loss function is defined as follows:
\begin{align}
\nonumber
\mathcal{L} = & \eta_{ID}^{Pix} \mathcal{L}_{ID}^{Pix} + \eta_{ID}^{Per} \mathcal{L}_{ID}^{Per} + \eta_{ID}^{Adv} \mathcal{L}_{ID}^{Adv} \\
& +\eta_{DD}^{Pix} \mathcal{L}_{DD}^{Pix} + \eta_{DD}^{Adv} \mathcal{L}_{DD}^{Adv}.
 \label{eq:loss}
\end{align}
The hyper-parameter $\eta$ leverages the penalties on different losses.

\newcommand{\tabincell}[2]{\begin{tabular}{@{}#1@{}}#2\end{tabular}}
\begin{table*}[t]\small
\caption{Ablation study. The best and second best results are shown in \textcolor{red}{red} and \textcolor{blue}{blue}.}
\label{tab:ablation}
    \centering
    \begin{tabular}{|c|ccc|c|c|c|c|c|c|c|c|c|}
    \hline
        \multirow{2}{*} {\textbf{No.}} & \multirow{2}{*} {\bm{$\mathcal{L}_{DD}^{Pix}$}} & \multirow{2}{*} {\bm{$\mathcal{L}_{DD}^{Adv}$}} & \multirow{2}{*} {\textbf{Attention}} & \multicolumn{3}{c|} {\bf{Set5}} & \multicolumn{3}{c|} {\bf{Set14}} & \multicolumn{3}{c|} {\bf{B100}} \\
        \cline{5-13}
          & & & & \textbf{PI} & \textbf{LPIPS} & \textbf{PSNR} & \textbf{PI} & \textbf{LPIPS} & \textbf{PSNR} &  \textbf{PI} & \textbf{LPIPS} & \textbf{PSNR} \\
          \hline
          \hline
        1 & \xmark & \xmark & \xmark & 3.4177 & 0.0656 & \textcolor[rgb]{ 0,  0,  1}{30.232} & 2.9016 & 0.1330 & \textcolor[rgb]{ 1,  0,  0}{26.566} & 2.3994 & 0.1637 & 25.396 \\
        2 & \cmark & \xmark & \xmark & \textcolor[rgb]{ 0,  0,  1}{3.2641} & 0.0685 & 30.161 & 2.8334 & 0.1369 & 26.453 & 2.3975 & 0.1615 & 25.336 \\
        3 & \cmark & \cmark & \xmark & 3.4642 & \textcolor[rgb]{ 1,  0,  0}{0.0640} & \textcolor[rgb]{ 1,  0,  0}{30.301} & \textcolor[rgb]{ 0,  0,  1}{2.7814} & \textcolor[rgb]{ 0,  0,  1}{0.1324} & 26.504 & \textcolor[rgb]{ 0,  0,  1}{2.3543} & \textcolor[rgb]{ 0,  0,  1}{0.1596} & \textcolor[rgb]{ 0,  0,  1}{25.452} \\
        4 & \cmark & \cmark & \cmark & \textcolor[rgb]{ 1,  0,  0}{3.2019} & \textcolor[rgb]{ 0,  0,  1}{0.0641} & 30.172 & \textcolor[rgb]{ 1,  0,  0}{2.7387} & \textcolor[rgb]{ 1,  0,  0}{0.1303} & \textcolor[rgb]{ 0,  0,  1}{26.554} & \textcolor[rgb]{ 1,  0,  0}{2.3525} & \textcolor[rgb]{ 1,  0,  0}{0.1590} & \textcolor[rgb]{ 1,  0,  0}{25.645} \\ \hline
    \end{tabular}
\end{table*}

\subsection{Differences between the DSRGAN and SPSR}
The proposed DSRGAN shares a similar network architecture to the SPSR in \cite{ma2020structure}. However, the DSRGAN is fundamentally different from the SPSR: a) The SPSR includes a gradient branch to learn image gradients, while the DSRGAN focus on image details. Compared with the gradients, the details are able to record fine textures in addition to sharp edges. Thus, the DSRGAN can pay more attention on recovering these high frequency information; b) In the DSRGAN, the fusion of SISR branch and detail branch does not rely on feature concatenation as in the SPSR. The fusion is realized by a more interpretable way, where the detail layer is multiplied back to the image domain. Such a fusion manner adheres to the conventional detail enhancement algorithm \cite{wls} and brings a more powerful ability to the network in generating high perceptual quality images; c) The gradient extraction has to be differentiable for training the SPSR. While in the DSRGAN, the detail extraction $\mathcal{E}$ is not required to be differentiable for an end-to-end training. Therefore, following the DSRGAN, it is feasible to blend other model-based conventional algorithms with recent data-driven deep learning methods and take advantages from both of them.

Notably, the SISR and detail branches in the DSRGAN are agnostic to the specific selection of network architectures, and any other deep learning models are feasible. Therefore, the DSRGAN is able to benefit from off-the-shelf networks that have powerful representation abilities.

\section{Experiments}
Extensive experiments are conducted to demonstrate the performance of our proposed DSRGAN. For qualitative comparison, readers are invited to view the electronic version of figures and zoom in them so as to better check differences among all images.

\subsection{Datasets and Metrics}
\subsubsection{Datasets}
The training split of DIV2K \cite{ntire} is used for training. It contains 800 2k-resolution images. For test, 5 commonly used datasets including Set5, Set14, BSD100, Urban100, and General100 are used. For both training and test, the LR images are stimulated by downsampling the HR images with the scaling factor as 4 via Bicubic kernel.
\subsubsection{Metrics}
For evaluating the quality of SR images, two metrics on measuring fidelity and two metrics on measuring perceptual quality are used, respectively. The fidelity metrics include PSNR and SSIM, which compute the dissimilarities between SR and HR images. One of the perceptual metrics, LPIPS \cite{zhang2018unreasonable}, computes the dissimilarities on feature-domain. The other one, PI \cite{pirm}, is a no-reference metric which does not need the ground truth HR images for evaluation. For the PSNR and SSIM, higher is better. For the PI and LPIPS, lower is better.

\subsection{Implementation details}
We use Pytorch based on Python to implement all methods. Experiments are conducted on the 4029GP-TRT server carried with NVIDIA TITAN RTX GPU of 24GB memory. All the metrics of SR results are ccomputed using MATLAB 2020a. We give more details about the method implementation and network training.
\subsubsection{Method details}
For the proposed detail extraction algorithm, $\alpha$ in Equations (\ref{eq:vector1}) and (\ref{eq:vector2}) is selected as 4 to amplify the details. $\lambda$ in Equation (\ref{eq:quadoptimize}) is selected as 1 for balancing the regularization and fidelity terms, such that the extracted details can be sparse.
$\gamma$ and $\epsilon$ in Equation (\ref{eq:psi}) are selected as 0.75 and 2 for simultaneously preserving details and excluding noises. The iteration times $T$ in Equation (\ref{eq:approximate}) is selected as 4.
For the generator network, the number of feature channels of the RRDB in the SISR branch is selected as 64 \cite{esrgan}. While the number of feature channels of the RRDB in the detail branch is selected as 128 \cite{ma2020structure}.
\subsubsection{Training details}
Before training the DSRGAN, we first use the pre-trained fidelity-oriented model in \cite{esrgan} to initialize the network parameters of the main SISR branch. During training, the HR and corresponding LR images are randomly cropped into patches with the size of $128 \times 128$ and $32 \times 32$ on the fly. The batch size is set as 20. The random cropping, flipping, and rotating are applied for data augmentation. In the loss function (\ref{eq:loss}), $\eta_{ID}^{Pix}$, $\eta_{ID}^{Per}$, and $\eta_{DD}^{Pix}$ are set as 0.2, 1, and 0.5, respectively. The $\eta_{ID}^{Adv}$ for the generator and discriminator are set as 0.005 and 1, respectively, the same for $\eta_{DD}^{Adv}$. The optimizer is Adam \cite{adam} with $\beta_1=0.9$ and $\beta_2=0.999$. The initial learning rates for generator and two discriminators are all set as 1e-4, and they are decreased by a factor of 0.5 at 50k, 100k, 200k, and 300k iterations. The number of total training iterations is up to 400k and it takes about 4-5 days to finish the training.

\begin{figure}[t] 
   {
    \begin{minipage}[b]{1\linewidth} 
      \centering
      \includegraphics[width=1.95cm,height=1.9cm]{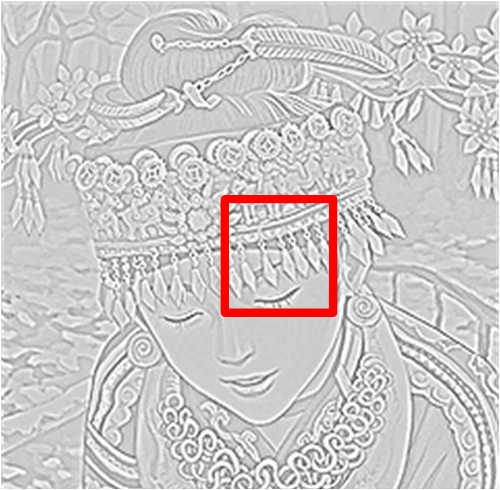}\vspace{-1pt}
      \includegraphics[width=1.95cm,height=1.9cm]{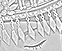}\vspace{-1pt}
      \includegraphics[width=1.95cm,height=1.9cm]{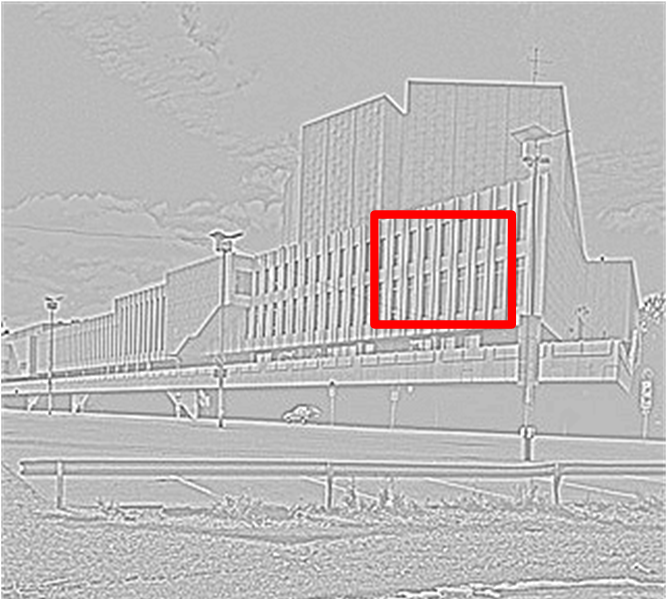}\vspace{-1pt}
      \includegraphics[width=1.95cm,height=1.9cm]{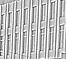}\vspace{-1pt}
    \end{minipage}
  }
  \centerline{\footnotesize{(a) Ground truth HR details}}\vspace{2pt}
   {
    \begin{minipage}[b]{1\linewidth} 
      \centering
      \includegraphics[width=1.95cm,height=1.9cm]{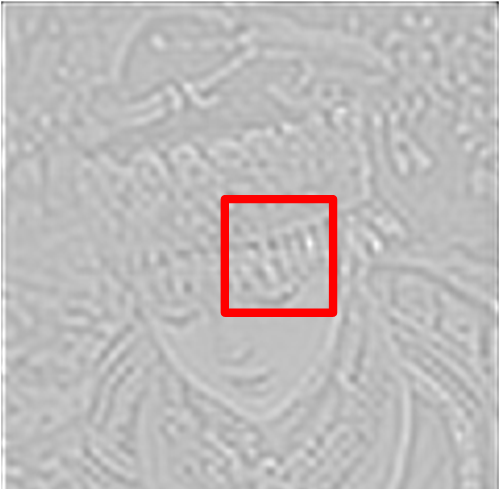}\vspace{-1pt}
      \includegraphics[width=1.95cm,height=1.9cm]{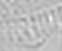}\vspace{-1pt}
      \includegraphics[width=1.95cm,height=1.9cm]{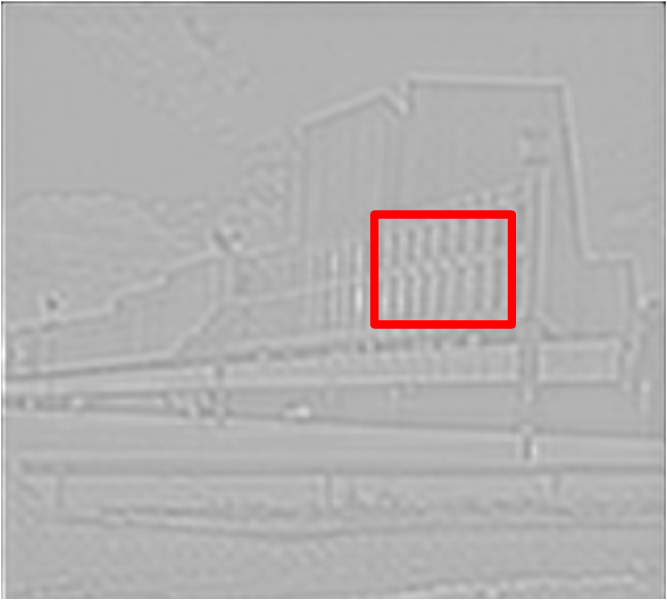}\vspace{-1pt}
      \includegraphics[width=1.95cm,height=1.9cm]{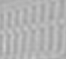}\vspace{-1pt}
    \end{minipage}
  }
  \centerline{\footnotesize{(b) Details supervised by $\mathcal{L}_{DD}^{Pix}$}}\vspace{2pt}
     {
    \begin{minipage}[b]{1\linewidth} 
      \centering
      \includegraphics[width=1.95cm,height=1.9cm]{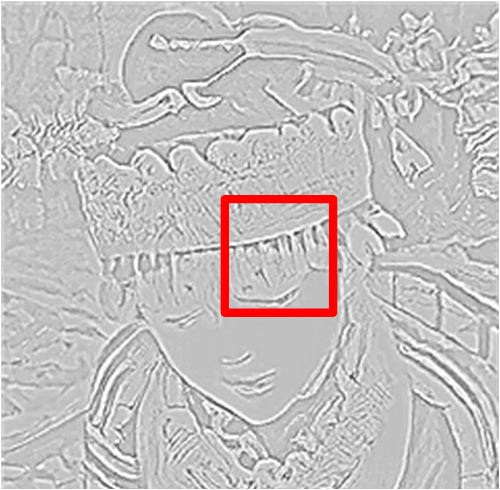}\vspace{-1pt}
      \includegraphics[width=1.95cm,height=1.9cm]{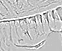}\vspace{-1pt}
      \includegraphics[width=1.95cm,height=1.9cm]{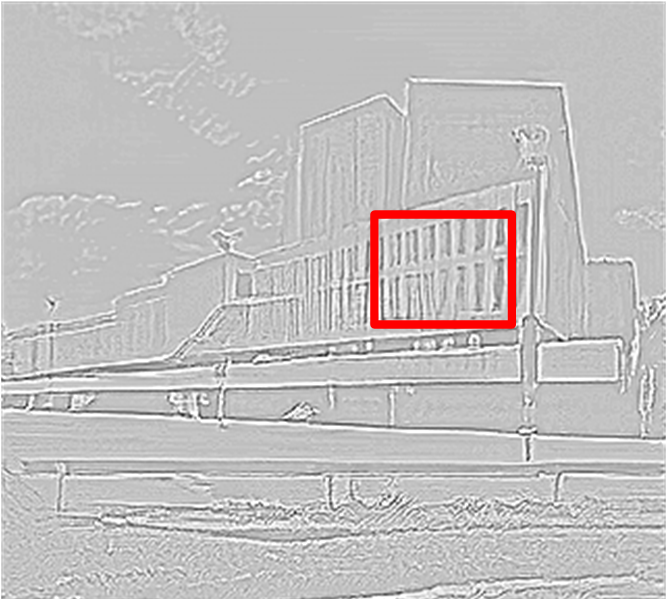}\vspace{-1pt}
      \includegraphics[width=1.95cm,height=1.9cm]{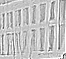}\vspace{-1pt}
    \end{minipage}
  }
  \centerline{\footnotesize{(c) Details supervised by $\mathcal{L}_{DD}^{Pix}$ and $\mathcal{L}_{DD}^{Adv}$}}\vspace{2pt}
     {
    \begin{minipage}[b]{1\linewidth} 
      \centering
      \includegraphics[width=1.95cm,height=1.9cm]{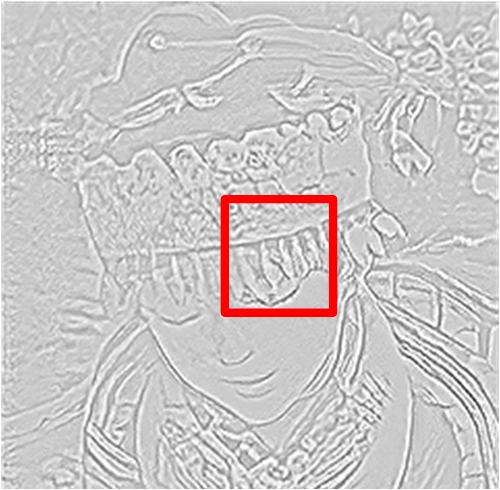}\vspace{-1pt}
      \includegraphics[width=1.95cm,height=1.9cm]{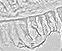}\vspace{-1pt}
      \includegraphics[width=1.95cm,height=1.9cm]{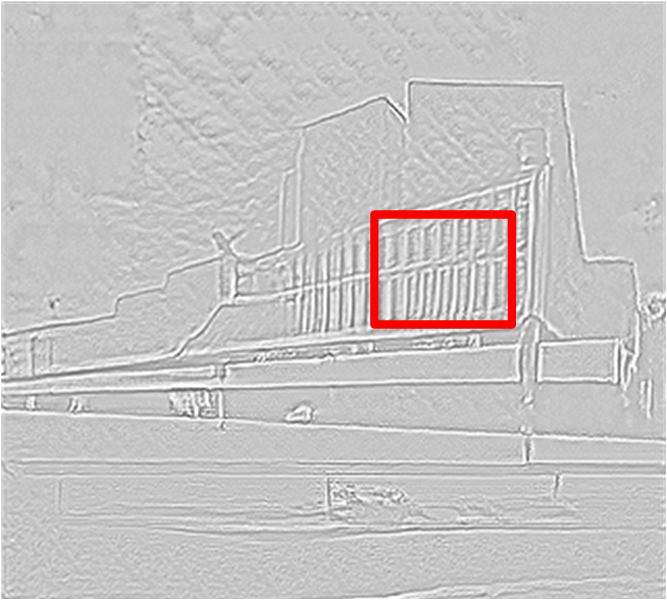}\vspace{-1pt}
      \includegraphics[width=1.95cm,height=1.9cm]{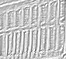}\vspace{-1pt}
    \end{minipage}
  }
  \centerline{\footnotesize{(d) Details supervised by $\mathcal{L}_{DD}^{Pix}$ and $\mathcal{L}_{DD}^{Adv}$ and refined by attention}}\vspace{2pt}
  \vfill
  \caption{The visualization of details. They are normalized into the range from 0 to 1. The details are exhibited on the left column and their corresponding parts in the red rectangular box are enlarged on the right. The examples are `comic' from Set14 and `78004' from B100.
 }
 \label{fig:ablation}
\end{figure}

\begin{table*}[t]\small
  \centering
  \caption{Comparison with state-of-the-art perception-oriented SISR methods. The best and second best results are shown in \textcolor{red}{red} and \textcolor{blue}{blue}.}
    \begin{tabular}{|l|c|c|c|c|c|c|c|c|c|c|}
    \hline
    \multicolumn{1}{|l|}{\multirow{2}[4]{*}{\textbf{Method}}} & \multicolumn{2}{c|}{\textbf{Set5}} & \multicolumn{2}{c|}{\textbf{Set14}} & \multicolumn{2}{c|}{\textbf{B100}} & \multicolumn{2}{c|}{\textbf{Urban100}} & \multicolumn{2}{c|}{\textbf{General100}} \\
\cline{2-11}         & \textbf{PSNR} & \textbf{SSIM} & \textbf{PSNR} & \textbf{SSIM} & \textbf{PSNR} & \textbf{SSIM} & \textbf{PSNR} & \textbf{SSIM} & \textbf{PSNR} & \textbf{SSIM} \\
    \hline
    \hline
    Bicubic & 28.386 & 0.8249 & 26.097 & 0.7853 & 25.961 & 0.6675 & 23.145 & 0.9011 & 28.018 & 0.8281 \\
    SRGAN \cite{srgan} (CVPR'17) & 29.879 & \textcolor[rgb]{ 1,  0,  0}{0.8694} & 26.552 & 0.7891 & 25.498 & 0.6520 & 24.387 & 0.9376 & 29.358 & 0.8531 \\
    EnhanceNet \cite{enhancenet} (ICCV'17) & 28.548 & 0.8374 & 25.763 & 0.7737 & 24.931 & 0.6259 & 23.543 & 0.9357 & 28.069 & 0.8308 \\
    SFTGAN \cite{sftgan} (CVPR'18) & 29.913 & 0.8672 & 26.226 & 0.7860 & 25.505 & 0.6549 & 24.015 & 0.9365 & 29.034 & 0.8512 \\
    ESRGAN \cite{esrgan} (ECCV'18) & \textcolor[rgb]{ 1,  0,  0}{30.422} & \textcolor[rgb]{ 0, 0, 1}{0.8683} & 26.272 & 0.7837 & 25.317 & 0.6506 & 24.360 & 0.9452 & 29.412 & 0.8546 \\
    SinGAN \cite{singan} (ICCV'19) & 26.419 & 0.7920 & 23.134 & 0.6807 & 24.330 & 0.5950 & 21.300 & 0.8203 & 25.392 & 0.7664 \\
    SPSR \cite{ma2020structure} (CVPR'20) & \textcolor[rgb]{ 0, 0, 1}{30.382} & 0.8635 & \textcolor[rgb]{ 0, 0, 1}{26.635} & \textcolor[rgb]{ 0, 0, 1}{0.7939} & 25.505 & 0.6576 & \textcolor[rgb]{ 0, 0, 1}{24.799} & 0.9481 & 29.414 & 0.8536 \\
    SRFlow \cite{srflow} (ECCV'20) & 30.184 & 0.8609 & \textcolor[rgb]{ 1,  0,  0}{26.880} & \textcolor[rgb]{ 1,  0,  0}{0.8006} & \textcolor[rgb]{ 1,  0,  0}{26.055} & \textcolor[rgb]{ 1,  0,  0}{0.6721} & \textcolor[rgb]{ 1,  0,  0}{25.268} & \textcolor[rgb]{ 1,  0,  0}{0.9535} & \textcolor[rgb]{ 1,  0,  0}{29.770} & \textcolor[rgb]{ 1,  0,  0}{0.8591} \\
    RankSRGAN \cite{ranksrgan} (TPAMI'21) & 29.640 & 0.8587 & 26.446 & 0.7864 & 25.453 & 0.6484 & 24.472 & 0.9413 & 29.098 & 0.8480 \\
    Fourier \cite{Fuoli} (ICCV'21) & -     & -     & -     & -     & 25.660 & 0.6560 & 24.690 & 0.7230 & -     & - \\
    HCFlow \cite{hcflow} (ICCV'21) & 30.360 & 0.8636 & 26.146 & 0.7809 & 25.537 & 0.6608 & 24.510 & 0.9471 & \textcolor[rgb]{ 0, 0, 1}{29.646} & \textcolor[rgb]{ 0, 0, 1}{0.8557} \\
    DSRGAN (Ours) & 30.172 & 0.8607 & 26.554 & 0.7932 & \textcolor[rgb]{ 0, 0, 1}{25.645} & \textcolor[rgb]{ 0, 0, 1}{0.6614} & 24.686 & \textcolor[rgb]{ 0, 0, 1}{0.9482} & 29.476 & 0.8548 \\
    \hline
    \hline
          & \textbf{PI} & \textbf{LPIPS} & \textbf{PI} & \textbf{LPIPS} & \textbf{PI} & \textbf{LPIPS} & \textbf{PI} & \textbf{LPIPS} & \textbf{PI} & \textbf{LPIPS} \\
    \hline
    \hline
    Bicubic & 7.3903 & 0.3440 & 7.0207 & 0.4412 & 7.0010 & 0.5249 & 6.9413 & 0.4726 & 7.9351 & 0.3528 \\
    SRGAN \cite{srgan} (CVPR'17) & 3.5684 & 0.0814 & 2.9234 & 0.1452 & 2.3816 & 0.1779 & 3.4848 & 0.1425 & 4.2173 & 0.0961 \\
    EnhanceNet \cite{enhancenet} (ICCV'17) & \textcolor[rgb]{ 1,  0,  0}{2.9600} & 0.1015 & 2.9842 & 0.1629 & 2.9071 & 0.2013 & \textcolor[rgb]{ 0, 0, 1}{3.4649} & 0.1632 & 4.1251 & 0.1327 \\
    SFTGAN \cite{sftgan} (CVPR'18) & 3.7726 & 0.0892 & 2.9027 & 0.1489 & 2.3775 & 0.1769 & 3.6141 & 0.1433 & 4.2843 & 0.1028 \\
    ESRGAN \cite{esrgan} (ECCV'18) & 3.7760 & 0.0758 & 2.8780 & 0.1332 & 2.4806 & 0.1614 & 3.7628 & 0.1229 & 4.3188 & 0.0879 \\
    SinGAN \cite{singan} (ICCV'19) & 4.1939 & 0.1901 & 2.9864 & 0.3034 & 2.6490 & 0.2916 & 3.7266 & 0.2657 & 4.1062 & 0.2208 \\
    SPSR \cite{ma2020structure} (CVPR'20) & 3.3124 & \textcolor[rgb]{ 0, 0, 1}{0.0647} & 2.8770 & 0.1327 & \textcolor[rgb]{ 0, 0, 1}{2.3513} & \textcolor[rgb]{ 0, 0, 1}{0.1610} & 3.5437 & \textcolor[rgb]{ 0, 0, 1}{0.1184} & 4.0969 & 0.0863 \\
    SRFlow \cite{srflow} (ECCV'20) & 4.2089 & 0.0818 & 3.0433 & \textcolor[rgb]{ 0, 0, 1}{0.1316} & 2.5971 & 0.1828 & 3.7927 & 0.1269 & 4.6271 & 0.0956 \\
    RankSRGAN \cite{ranksrgan} (TPAMI'21) & \textcolor[rgb]{ 0, 0, 1}{3.1266} & 0.0734 & \textcolor[rgb]{ 1,  0,  0}{2.5123} & 0.1371 & \textcolor[rgb]{ 1,  0,  0}{2.1234} & 0.1745 & \textcolor[rgb]{ 1,  0,  0}{3.3304} & 0.1380 & \textcolor[rgb]{ 1,  0,  0}{3.8916} & 0.0955 \\
    Fourier \cite{Fuoli} (ICCV'21) & -     & -     & -     & -     & -     & 0.1720 & -     & 0.1320 & -     & - \\
    HCFlow \cite{hcflow} (ICCV'21) & 4.0203 & 0.0752 & 2.9530 & 0.1325 & 2.5951 & 0.1638 & 3.6320 & 0.1242 & 4.3690 & \textcolor[rgb]{ 0, 0, 1}{0.0859} \\
    DSRGAN (Ours) & 3.2019 & \textcolor[rgb]{ 1,  0,  0}{0.0641} & \textcolor[rgb]{ 0, 0, 1}{2.7387} & \textcolor[rgb]{ 1,  0,  0}{0.1303} & 2.3525 & \textcolor[rgb]{ 1,  0,  0}{0.1590} & {3.4918} & \textcolor[rgb]{ 1,  0,  0}{0.1175} & \textcolor[rgb]{ 0, 0, 1}{4.0319} & \textcolor[rgb]{ 1,  0,  0}{0.0847} \\
    \hline
    \end{tabular}%
  \label{tab:sota}%
\end{table*}%

\begin{figure*}[t]
  \centering
  \begin{minipage}[b]{0.97\linewidth}
   {
    \begin{minipage}[b]{0.22\linewidth}
      \centering
      \includegraphics[width=3.8cm,height=3.8cm]{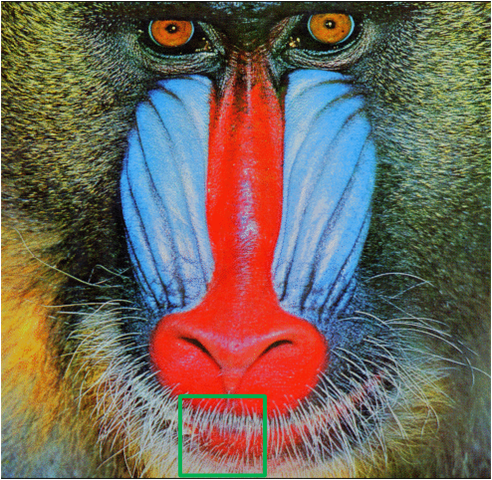}
      \centerline{\footnotesize{`baboon' from Set14}}\vspace{5pt}
      \includegraphics[width=3.8cm,height=3.8cm]{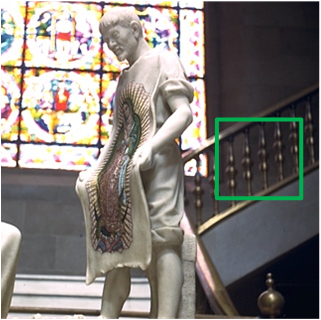}\vspace{0pt}
      \centerline{\footnotesize{`24077' from B100}}\vspace{5pt}
      \includegraphics[width=3.8cm,height=3.8cm]{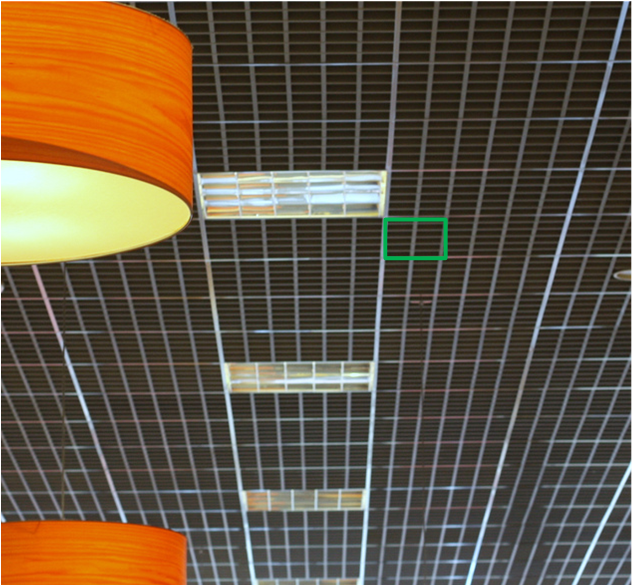}\vspace{0pt}
      \centerline{\footnotesize{`img\_044' from Urban100}}\vspace{5pt}
      \includegraphics[width=3.8cm,height=3.8cm]{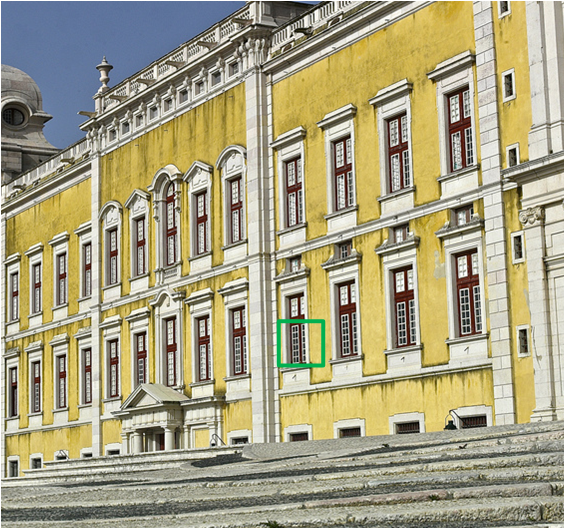}\vspace{0pt}
      \centerline{\footnotesize{`img\_054' from Urban100}}
    \end{minipage}
  }
   {
    \begin{minipage}[b]{0.11\linewidth}
      \centering
      \includegraphics[width=1.95cm,height=1.6cm]{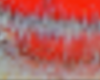}\vspace{0pt}
      \centerline{\footnotesize{Bicubic}}\vspace{5pt}
      \includegraphics[width=1.95cm,height=1.6cm]{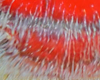}\vspace{0pt}
      \centerline{\footnotesize{SPSR}}\vspace{5pt}
      \includegraphics[width=1.95cm,height=1.6cm]{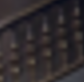}\vspace{0pt}
      \centerline{\footnotesize{Bicubic}}\vspace{5pt}
      \includegraphics[width=1.95cm,height=1.6cm]{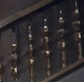}\vspace{0pt}
      \centerline{\footnotesize{SPSR}}\vspace{5pt}
      \includegraphics[width=1.95cm,height=1.6cm]{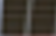}\vspace{0pt}
      \centerline{\footnotesize{Bicubic}}\vspace{5pt}
      \includegraphics[width=1.95cm,height=1.6cm]{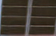}\vspace{0.6pt}
      \centerline{\footnotesize{SPSR}}\vspace{6pt}
      \includegraphics[width=1.95cm,height=1.6cm]{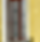}\vspace{0pt}
      \centerline{\footnotesize{Bicubic}}\vspace{5pt}
      \includegraphics[width=1.95cm,height=1.6cm]{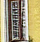}\vspace{0pt}
      \centerline{\footnotesize{SPSR}}

    \end{minipage}
  }
     {
    \begin{minipage}[b]{0.11\linewidth}
      \centering
      \includegraphics[width=1.95cm,height=1.6cm]{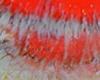}\vspace{0pt}
      \centerline{\footnotesize{SRGAN}}\vspace{5pt}
      \includegraphics[width=1.95cm,height=1.6cm]{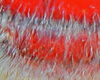}\vspace{0pt}
      \centerline{\footnotesize{SRFlow}}\vspace{5pt}
      \includegraphics[width=1.95cm,height=1.6cm]{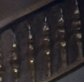}\vspace{0pt}
      \centerline{\footnotesize{SRGAN}}\vspace{5pt}
      \includegraphics[width=1.95cm,height=1.6cm]{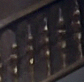}\vspace{0pt}
      \centerline{\footnotesize{SRFlow}}\vspace{5pt}
      \includegraphics[width=1.95cm,height=1.6cm]{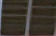}\vspace{0pt}
      \centerline{\footnotesize{SRGAN}}\vspace{5pt}
      \includegraphics[width=1.95cm,height=1.6cm]{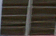}\vspace{0.6pt}
      \centerline{\footnotesize{SRFlow}}\vspace{6pt}
      \includegraphics[width=1.95cm,height=1.6cm]{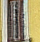}\vspace{0pt}
      \centerline{\footnotesize{SRGAN}}\vspace{5pt}
      \includegraphics[width=1.95cm,height=1.6cm]{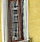}\vspace{0pt}
      \centerline{\footnotesize{SRFlow}}

    \end{minipage}
  }
     {
    \begin{minipage}[b]{0.11\linewidth}
      \centering
      \includegraphics[width=1.95cm,height=1.6cm]{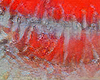}\vspace{0pt}
      \centerline{\footnotesize{EnhanceNet}}\vspace{5pt}
      \includegraphics[width=1.95cm,height=1.6cm]{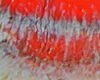}\vspace{0pt}
      \centerline{\footnotesize{RankSRGAN}}\vspace{5pt}
      \includegraphics[width=1.95cm,height=1.6cm]{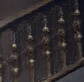}\vspace{0pt}
      \centerline{\footnotesize{EnhanceNet}}\vspace{5pt}
      \includegraphics[width=1.95cm,height=1.6cm]{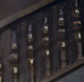}\vspace{0pt}
      \centerline{\footnotesize{RankSRGAN}}\vspace{5pt}
      \includegraphics[width=1.95cm,height=1.6cm]{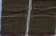}\vspace{0pt}
      \centerline{\footnotesize{EnhanceNet}}\vspace{5pt}
      \includegraphics[width=1.95cm,height=1.6cm]{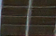}\vspace{0.6pt}
      \centerline{\footnotesize{RankSRGAN}}\vspace{6pt}
      \includegraphics[width=1.95cm,height=1.6cm]{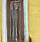}\vspace{0pt}
      \centerline{\footnotesize{EnhanceNet}}\vspace{5pt}
      \includegraphics[width=1.95cm,height=1.6cm]{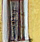}\vspace{0pt}
      \centerline{\footnotesize{RankSRGAN}}

    \end{minipage}
  }
     {
    \begin{minipage}[b]{0.11\linewidth}
      \centering
      \includegraphics[width=1.95cm,height=1.6cm]{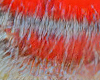}\vspace{0pt}
      \centerline{\footnotesize{SFTGAN}}\vspace{5pt}
      \includegraphics[width=1.95cm,height=1.6cm]{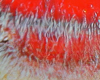}\vspace{0pt}
      \centerline{\footnotesize{HCFlow}}\vspace{5pt}
      \includegraphics[width=1.95cm,height=1.6cm]{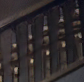}\vspace{0pt}
      \centerline{\footnotesize{SFTGAN}}\vspace{5pt}
      \includegraphics[width=1.95cm,height=1.6cm]{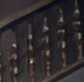}\vspace{0pt}
      \centerline{\footnotesize{HCFlow}}\vspace{5pt}
      \includegraphics[width=1.95cm,height=1.6cm]{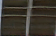}\vspace{0pt}
      \centerline{\footnotesize{SFTGAN}}\vspace{5pt}
      \includegraphics[width=1.95cm,height=1.6cm]{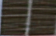}\vspace{0.6pt}
      \centerline{\footnotesize{HCFlow}}\vspace{6pt}
      \includegraphics[width=1.95cm,height=1.6cm]{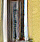}\vspace{0pt}
      \centerline{\footnotesize{SFTGAN}}\vspace{5pt}
      \includegraphics[width=1.95cm,height=1.6cm]{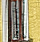}\vspace{0pt}
      \centerline{\footnotesize{HCFlow}}

    \end{minipage}
  }
     {
    \begin{minipage}[b]{0.11\linewidth}
      \centering
      \includegraphics[width=1.95cm,height=1.6cm]{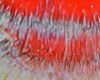}\vspace{0pt}
      \centerline{\footnotesize{ESRGAN}}\vspace{5pt}
      \includegraphics[width=1.95cm,height=1.6cm]{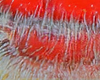}\vspace{0pt}
      \centerline{\footnotesize{DSRGAN}}\vspace{5pt}
      \includegraphics[width=1.95cm,height=1.6cm]{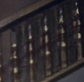}\vspace{0pt}
      \centerline{\footnotesize{ESRGAN}}\vspace{5pt}
      \includegraphics[width=1.95cm,height=1.6cm]{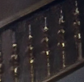}\vspace{0pt}
      \centerline{\footnotesize{DSRGAN}}\vspace{5pt}
      \includegraphics[width=1.95cm,height=1.6cm]{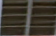}\vspace{0pt}
      \centerline{\footnotesize{ESRGAN}}\vspace{5pt}
      \includegraphics[width=1.95cm,height=1.6cm]{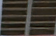}\vspace{0.6pt}
      \centerline{\footnotesize{DSRGAN}}\vspace{6pt}
      \includegraphics[width=1.95cm,height=1.6cm]{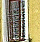}\vspace{0pt}
      \centerline{\footnotesize{ESRGAN}}\vspace{5pt}
      \includegraphics[width=1.95cm,height=1.6cm]{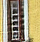}\vspace{0pt}
      \centerline{\footnotesize{DSRGAN}}

    \end{minipage}
  }
     {
    \begin{minipage}[b]{0.11\linewidth}
      \centering
      \includegraphics[width=1.95cm,height=1.6cm]{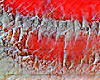}\vspace{0pt}
      \centerline{\footnotesize{SinGAN}}\vspace{5pt}
      \includegraphics[width=1.95cm,height=1.6cm]{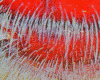}\vspace{0pt}
      \centerline{\footnotesize{Ground Truth}}\vspace{5pt}
      \includegraphics[width=1.95cm,height=1.6cm]{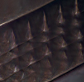}\vspace{0pt}
      \centerline{\footnotesize{SinGAN}}\vspace{5pt}
      \includegraphics[width=1.95cm,height=1.6cm]{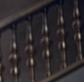}\vspace{0pt}
      \centerline{\footnotesize{Ground Truth}}\vspace{5pt}
      \includegraphics[width=1.95cm,height=1.6cm]{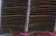}\vspace{0pt}
      \centerline{\footnotesize{SinGAN}}\vspace{5pt}
      \includegraphics[width=1.95cm,height=1.6cm]{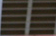}\vspace{0.6pt}
      \centerline{\footnotesize{Ground Truth}}\vspace{6pt}
      \includegraphics[width=1.95cm,height=1.6cm]{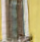}\vspace{0pt}
      \centerline{\footnotesize{SinGAN}}\vspace{5pt}
      \includegraphics[width=1.95cm,height=1.6cm]{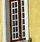}\vspace{0pt}
      \centerline{\footnotesize{Ground Truth}}

    \end{minipage}
  }

  \end{minipage}
  \vfill
  \caption{The qualitative comparison of SR images restored by state-of-the-art perception-oriented SISR methods. The original images are exhibited on the left column and their SR counterparts in the green rectangular box are enlarged on the right.
 }
 \label{fig:sota}
\end{figure*}

\subsection{Ablation study}
To investigate the effects of the model-based detail extraction algorithm on the data-driven deep learning network, the ablation study is conducted on Set5, Set14, and B100 datasets. The improvements are gradually added to the baseline method to see how they contribute to the SR results. Different ablation methods are listed on the left of Table \ref{tab:ablation}, and each of them is explained as follows:
\begin{itemize}
\item The No. 1 ablation represents the baseline method, where the detail extraction algorithm, the detail branch, and the detail-domain discriminator are disabled. In practice, it is implemented by an enhanced ESRGAN, whose generator is composed of 27 RRDBs instead of 23 for a fair comparison with the rest of ablation methods (i.e., for roughly equivalent number of network parameters).

\item The No. 2 ablation represents the DSRGAN casting off $\mathcal{L}_{DD}^{Adv}$ and the attention module, where the detail layer $ I_D^{LR}$ generated by the detail branch is only supervised by $\mathcal{L}_{DD}^{Pix}$ with penalty $\eta_{DD}^{Pix}$.

\item The No. 3 ablation represents the DSRGAN casting off the attention module, where $I_D^{LR}$ is supervised by $\mathcal{L}_{DD}^{Pix}$ with penalty $\eta_{DD}^{Pix}$ and by $\mathcal{L}_{DD}^{Adv}$ with penalty $\eta_{DD}^{Adv}$.

\item The No. 4 ablation represents the complete DSRGAN, where $I_D^{LR}$ is refined by the attention module before being multiplied back to the image domain.

\end{itemize}

According to Table \ref{tab:ablation}, the No. 1 and 2 ablation methods have their own strengths on three datasets in terms of LPIPS. The improvement is not significant as the $I_D^{SR}$ is restored badly by training with the $\mathcal{L}_{DD}^{Pix}$ only. As shown in Figures \ref{fig:ablation}(a) and \ref{fig:ablation}(b), edges and textures are seriously smoothed in the restored $I_D^{LR}$. This problem is the same with that met by fidelity-oriented SISR methods, where the L1 regularization is hard to capture high frequency differences.

The No. 3 ablation method improves the PI and LPIPS by a large margin from the No. 1 and 2. As shown in Figure \ref{fig:ablation}(c), the high frequency information is preserved well in $I_D^{SR}$ by adding the adversarial loss $\mathcal{L}_{DD}^{Adv}$. The experimental results demonstrate that an explicit learning of detail prior plays a very positive effect on the perceptual SISR reconstruction. The deep learning network is strengthened by the model-based conventional algorithm.

The refinement by attention module further highlights the edges and textures in $I_D^{SR}$, as shown in Figure \ref{fig:ablation}(d). The attention can also adaptively controls the degree of detail enhancement for per pixel. According to Table \ref{tab:ablation}, it improves the perceptual quality of SR images, though slightly. Notably, the PSNR is not affected by the detail enhancement and even gets better on a few datasets.

\subsection{Comparison with state-of-the-art methods}
The complete DSRGAN is compared with several state-of-the-art perception-oriented SISR methods on benchmark SR datasets. Notably, for a fair comparison, we implement all these methods (except for Fourier \cite{Fuoli}) by using the official source codes and trained model weights provided by their authors instead of retraining them from scratch. The experimental results of Fourier is directly quoted from its paper. All the results are summarized in Table \ref{tab:sota}.

For perceptual metrics, the proposed DSRGAN achieves the best LPIPS on all 5 datasets. The DSRGAN is also very competitive on the PI and achieves the second best on Set14 and General100. The RankSRGAN \cite{ranksrgan} performs the best PI on most datasets as it is trained in the direction of PI via a rank-content loss. However, its LPIPS is worse than most recent methods. Compared with the most relevant method SPSR \cite{ma2020structure}, the DSRGAN improves both PI and LPIPS significantly on all datasets. For the fidelity metrics, the DSRGAN still has an advantage over most perception-oriented SISR methods. It achieves the second best SSIM on B100 and Urban100. The SRFlow \cite{srflow} gets the highest PSNR and SSIM on most datasets. However, its performances on PI and LPIPS are very inferior to the DSRGAN. Since the SRFlow is a flow-based method and doesn't include the adversarial loss, it is more like a fidelity-oriented SISR method. The overall results demonstrate the superior ability of DSRGAN in producing nature and photo-realistic SR images with minor fidelity distortions simultaneously. In addition, the transcendence of the DSRGAN over the SFTGAN \cite{sftgan} and SPSR \cite{ma2020structure} indicates that the detail prior is a more powerful assistance for the perceptual SISR than the semantic or gradient priors. The detail prior partially alleviates the ill-posed problem of the SISR reconstruction.

A few restored SR images are shown in Figure \ref{fig:sota} for qualitative comparison. Readers are invited to view the electronic version of figures and zoom in them so as to better check differences  among all images.
For the `baboon' from Set14, the DSRGAN restores the beard as realistic as ground truth HR image.
This is benefiting from the explicit texture extraction by model-based conventional algorithm.
For the `24077' from B100, enabled by taking multiplication as the fusion manner, the DSRGAN generates even more clear and natural pillars than ground truth. For the `img\_044' and `img\_054' from Urban100, the SR images produced by DSRGAN have the most similar sharp edges to HR counterparts among all the methods. This shows that the detail prior is able to preserve sharp edges and structures, with the same as the gradient prior \cite{ma2020structure}. Notably, the detail prior additionally includes the fine texture information that the gradient prior doesn't have. Moreover, owing to the concept of detail enhancement \cite{wls}, the image restored by DSRGAN is even of higher perceptual quality than the original image. For instance, in the `img\_044', the luminance of edges produced by DSRGAN is even brighter than ground truth.


\begin{table}[t]\small
  \centering
  \caption{Comparison among DSRGANs assisted by different kinds of detail. The best results is shown in \textcolor{red}{red}.}
    \begin{tabular}{|l|c|c|c|c|}
    \hline
    \multicolumn{1}{|l|}{\multirow{2}[4]{*}{\textbf{Method}}} & \multicolumn{4}{c|}{\textbf{Set14}} \\
\cline{2-5}         & \textbf{PI} & \textbf{LPIPS} & \textbf{PSNR} & \textbf{SSIM} \\
    \hline
    \hline
        DSRGAN-GIF & 2.8215 & 0.1372 & 26.314 & 0.7871 \\
        DSRGAN-MSDM & 2.8484 & 0.1366 & 26.085 & 0.7884 \\
        DSRGAN & \textcolor[rgb]{ 1,  0,  0}{2.7387} & \textcolor[rgb]{ 1,  0,  0}{0.1303} & \textcolor[rgb]{ 1,  0,  0}{26.554} & \textcolor[rgb]{ 1,  0,  0}{0.7932} \\
    \hline
    \hline
    \multicolumn{1}{|l|}{\multirow{2}[4]{*}{\textbf{Method}}} & \multicolumn{4}{c|}{\textbf{B100}} \\
\cline{2-5}         & \textbf{PI} & \textbf{LPIPS} & \textbf{PSNR} & \textbf{SSIM} \\
    \hline
    \hline
        DSRGAN-GIF & \textcolor[rgb]{ 1,  0,  0}{2.3154} & 0.1602 & 25.543 & 0.6604 \\
        DSRGAN-MSDM & 2.4765 & 0.1606 & 25.066 & 0.6415 \\
        DSRGAN & 2.3525 & \textcolor[rgb]{ 1,  0,  0}{0.1590} & \textcolor[rgb]{ 1,  0,  0}{25.645} & \textcolor[rgb]{ 1,  0,  0}{0.6614} \\
    \hline
    \end{tabular}%
  \label{tab:detail}%
\end{table}%

\begin{figure}[htbp] 
   {
    \begin{minipage}[b]{1\linewidth} 
      \centering
      \includegraphics[width=3.8cm,height=2.3cm]{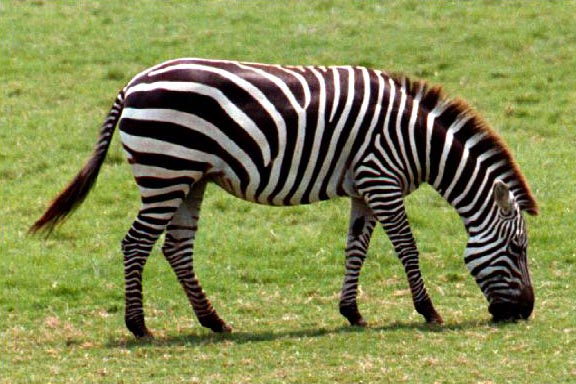}\vspace{-1pt}
      \includegraphics[width=3.8cm,height=2.3cm]{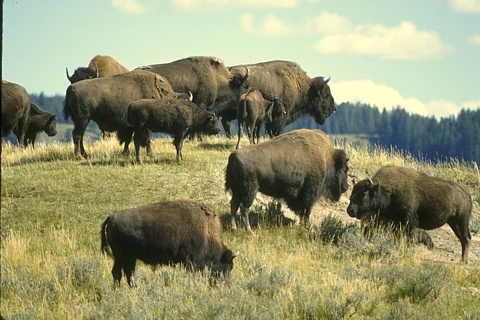}\vspace{-1pt}
    \end{minipage}
  }
  \centerline{\footnotesize{(a) Ground Truth HR images}}\vspace{2pt}
     {
    \begin{minipage}[b]{1\linewidth} 
      \centering
      \includegraphics[width=3.8cm,height=2.3cm]{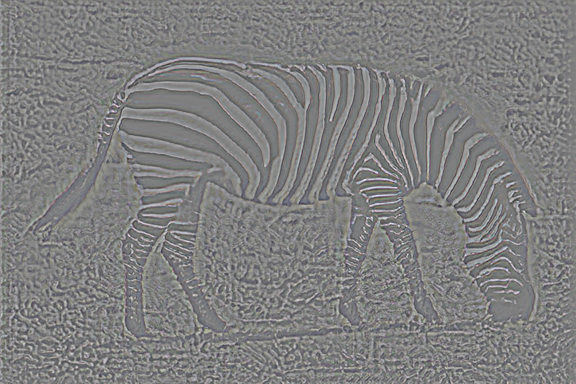}\vspace{-1pt}
      \includegraphics[width=3.8cm,height=2.3cm]{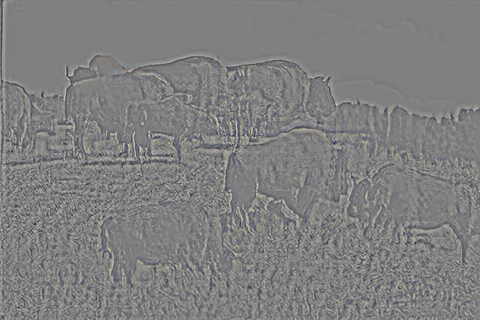}\vspace{-1pt}
    \end{minipage}
  }
  \centerline{\footnotesize{(b) Details produced by DSRGAN-GIF}}\vspace{2pt}
     {
    \begin{minipage}[b]{1\linewidth} 
      \centering
      \includegraphics[width=3.8cm,height=2.3cm]{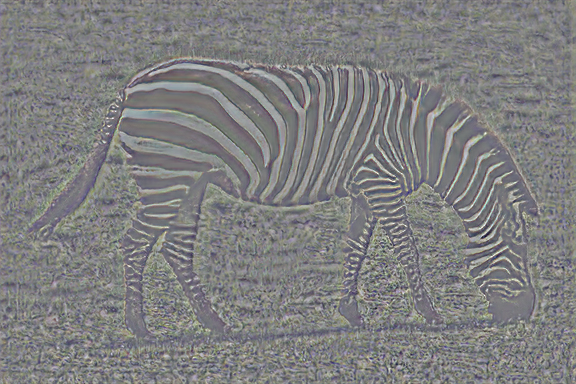}\vspace{-1pt}
      \includegraphics[width=3.8cm,height=2.3cm]{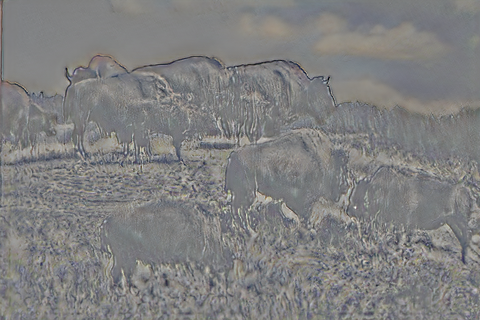}\vspace{-1pt}
    \end{minipage}
  }
  \centerline{\footnotesize{(c) Details produced by DSRGAN-MSDM}}\vspace{2pt}
     {
    \begin{minipage}[b]{1\linewidth}
      \centering
      \includegraphics[width=3.8cm,height=2.3cm]{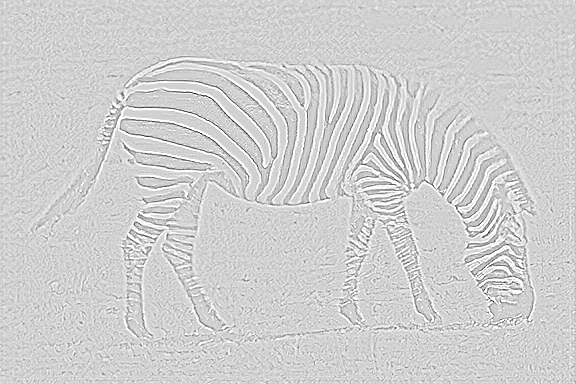}\vspace{-1pt}
      \includegraphics[width=3.8cm,height=2.3cm]{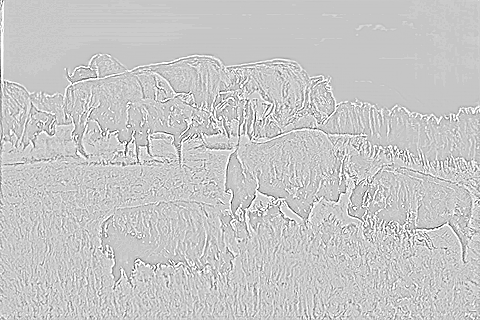}\vspace{-1pt}
    \end{minipage}
  }
  \centerline{\footnotesize{(d) Details produced by DSRGAN}}\vspace{2pt}
  \vfill
  \caption{The visualization of details. They are normalized into the range from 0 to 1. Notably, the details of DSRGAN have only one channel, while the details of DSRGAN-GIF and DSRGAN-MSDM have three channels. The examples are `zebra' from Set14 and `38092' from B100.
 }
 \label{fig:detail}
\end{figure}

\subsection{Analysis of the detail prior}
In this subsection, we replace $\mathcal{E}$ in the DSRGAN with other commonly used detail extraction algorithms to see how they affect the SISR reconstruction. In the field of image processing, $I_D$ is commonly extracted by subtracting $I_B$ from $I$, where $I_B$ is smoothed from $I$ by image smoothing algorithm. Two classic smoothing algorithms, guided image filtering (GIF) \cite{gif} and multi-scale detail manipulation (MSDM) \cite{farbman2008edge}, are utilized to replace the proposed detail extraction algorithm to obtain $I_D^{LR}$ and $I_D^{HR}$ in the complete DSRGAN.

For the GIF, the $r_{GIF}$ and $\epsilon_{GIF}$ in \cite{gif} are selected as 2 and 0.3. For the MSDM, the $\lambda_{MSDM}$ and $\alpha_{MSDM}$ in \cite{farbman2008edge} are selected as 1 and 1.2. We denote the new DSRGAN based on GIF and MSDM as DSRGAN-GIF and DSRGAN-MSDM, respectively. Notably, in the DSRGAN-GIF and DSRGAN-MSDM, the restored detail $I_D^{SR}$ is element-wise summed back to the image domain instead of multiplication, because the ground truth detail $I_D^{HR}$ is subtracted from $I^{HR}$.

The performances of DSRGAN, DSRGAN-GIF, and DSRGAN-MSDM are compared on Set14 and B100, as shown in Table \ref{tab:detail}. We can see that these methods achieve close PI and LPIPS, demonstrating that conventional detail extraction algorithms do have a positive effect on deep learning-based SISR. Therefore, combining model-based conventional algorithms with deep learning networks for the SISR reconstruction has a great potential research value. The combination can be achieved via the DSRGAN, where the conventional algorithms are not required to be differentiable.

The DSRGAN still has a slight advantage over the DSRGAN-GIF and DSRGAN-MSDM on both perceptual metrics and fidelity metrics. The reason might be that the detail extracted by the proposed algorithm is more sparse than those extracted by the GIF and MSDM. The sparse attribute originates from the regularization term in Equation (\ref{eq:quadoptimize}). As such, the burden of detail branch for restoring the $I_D^{SR}$ is greatly reduced. Another reason might be that image noises are filtered out from the $I_B$ by GIF and MSDM, but reappear in the $I_D$. The noises can be interferences to the image. We visualize the details restored by DSRGAN-GIF, DSRGAN-MSDM, and DSRGAN in Figure \ref{fig:detail}. The detail from DSRGAN do have fewer noises compared to the DSRGAN-GIF and DSRGAN-MSDM, and it is much more sparse. The details from DSRGAN-GIF and DSRGAN-MSDM not only preserve high frequency information, but also include unwanted redundant information such as colors and noises.

\begin{figure}[t]
  \centering
  \begin{minipage}[b]{1\linewidth}
  \centering

    \begin{minipage}[b]{0.32\linewidth}
      \centering
      \includegraphics[width=2.8cm,height=1.8cm]{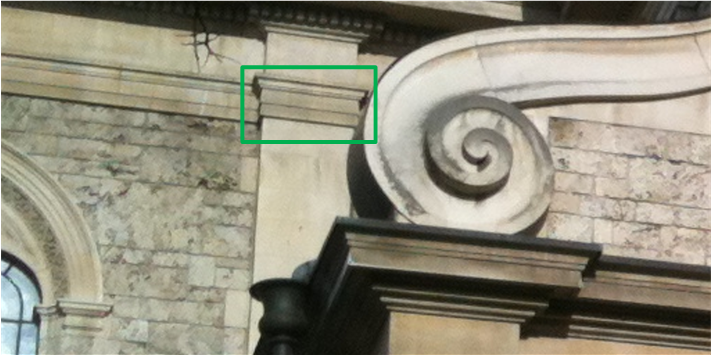}\vspace{3pt}
    \end{minipage}
    \begin{minipage}[b]{0.32\linewidth}
      \centering
      \includegraphics[width=2.8cm,height=1.8cm]{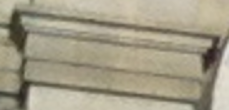}\vspace{3pt}
    \end{minipage}
        \begin{minipage}[b]{0.32\linewidth}
      \centering
      \includegraphics[width=2.8cm,height=1.8cm]{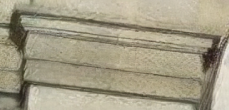}\vspace{3pt}
    \end{minipage}

    \end{minipage}

  \begin{minipage}[b]{1\linewidth}
  \centering

    \begin{minipage}[b]{0.32\linewidth}
      \centering
      \includegraphics[width=2.8cm,height=1.8cm]{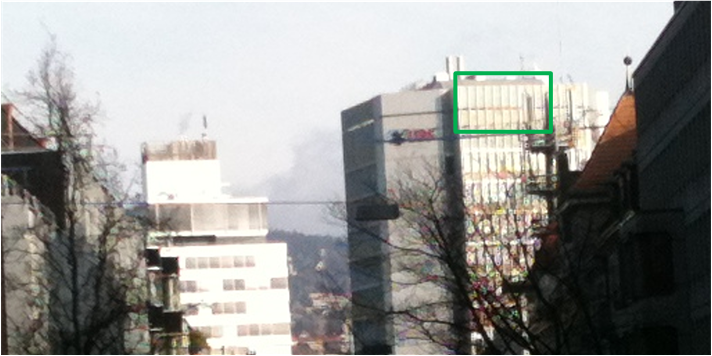}\vspace{-3pt}
      \centerline{\footnotesize{(a) LR images}}\vspace{0pt}
    \end{minipage}
    \begin{minipage}[b]{0.32\linewidth}
      \centering
      \includegraphics[width=2.8cm,height=1.8cm]{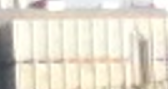}\vspace{-3pt}
      \centerline{\footnotesize{(b) Bicubic}}\vspace{0pt}
    \end{minipage}
        \begin{minipage}[b]{0.32\linewidth}
      \centering
      \includegraphics[width=2.8cm,height=1.8cm]{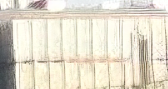}\vspace{-3pt}
      \centerline{\footnotesize{(c) DSRGAN}}\vspace{0pt}
    \end{minipage}

      \end{minipage}

  \vfill
  \caption{The performance of DSRGAN on real-world image SR. The original LR images are exhibited on the left column and their SR counterparts in the green rectangular box are enlarged on the right. The examples are `00047' and `00049'.}

 \label{fig:limitation}
\end{figure}

\subsection{Limitation of the proposed DSRGAN}
To investigate the generalization ability of the proposed DSRGAN, we conduct experiments on real-world image SR. The track 2 of NTIRE 2020 Challenge contains a set of LR images captured by smart phones, where ground truth HR images are non-existent \cite{lugmayr2020ntire}. The SR images restored by DSRGAN are shown in Figure \ref{fig:limitation}. It can be observed that the DSRGAN preserves edges clear for the real-world image SR. Nevertheless, on the second row, the SR image is carried with a few artifacts and noises, which demonstrates that there still exists a performance gap of DSRGAN between the benchmark SR and the real-world SR. The DSRGAN can be extended to the real-world image SR by taking advantages of real-world-oriented SR methods, such as \cite{ji2020real}. This will be our future work.

\section{Conclusion}
Inspired by conventional detail enhancement algorithms, a novel detail prior-assisted SISR paradigm via the GAN, named DSRGAN, is proposed in this study. Extensive experiments demonstrate that the proposed detail prior is a powerful assistance for the deep learning network in producing SR images with realistic details. The DSRGAN benefits from advantages of both conventional model-based algorithms and data-driven deep learning networks. It is also network-agnostic, which can be used for off-the-shelf SISR networks. Moreover, the detail extraction algorithm is not required to be differentiable. Following the DSRGAN, it is feasible to incorporate other conventional image processing algorithms into a deep learning network to form a model-based deep SISR, which will have a great potential research value.


One more interesting problem is to extend the DSRGAN to the video coding. Instead of coding a video directly, the video can be down-sampled. A super-resolution friendly rate control can be extended from existing rate control algorithms \cite{li2003adaptive, liu2006novel} to code the down-sampled video. Then, the video can be super-resolved by the extended DSRGAN. These problems will be studied in our future research.

\ifCLASSOPTIONcaptionsoff
  \newpage
\fi

\balance
\bibliographystyle{ieeetr}
\bibliography{mybib}

\begin{thebibliography}{10}

\bibitem{yang2010image}
J.~Yang, J.~Wright, T.~S. Huang, and Y.~Ma, ``Image super-resolution via sparse
  representation,'' {\em IEEE transactions on image processing}, vol.~19,
  no.~11, pp.~2861--2873, 2010.

\bibitem{pirm}
Y.~Blau, R.~Mechrez, R.~Timofte, T.~Michaeli, and L.~Zelnik-Manor, ``The 2018
  pirm challenge on perceptual image super-resolution,'' in {\em Proceedings of
  the European Conference on Computer Vision (ECCV)}, pp.~0--0, 2018.

\bibitem{zhang2018unreasonable}
R.~Zhang, P.~Isola, A.~A. Efros, E.~Shechtman, and O.~Wang, ``The unreasonable
  effectiveness of deep features as a perceptual metric,'' in {\em Proceedings
  of the IEEE conference on computer vision and pattern recognition},
  pp.~586--595, 2018.

\bibitem{srcnn}
C.~Dong, C.~C. Loy, K.~He, and X.~Tang, ``Image super-resolution using deep
  convolutional networks,'' {\em IEEE transactions on pattern analysis and
  machine intelligence}, vol.~38, no.~2, pp.~295--307, 2015.

\bibitem{vdsr}
J.~Kim, J.~Kwon~Lee, and K.~Mu~Lee, ``Accurate image super-resolution using
  very deep convolutional networks,'' in {\em Proceedings of the IEEE
  conference on computer vision and pattern recognition}, pp.~1646--1654, 2016.

\bibitem{edsr}
B.~Lim, S.~Son, H.~Kim, S.~Nah, and K.~Mu~Lee, ``Enhanced deep residual
  networks for single image super-resolution,'' in {\em Proceedings of the IEEE
  conference on computer vision and pattern recognition workshops},
  pp.~136--144, 2017.

\bibitem{wdsr}
J.~Yu, Y.~Fan, J.~Yang, N.~Xu, Z.~Wang, X.~Wang, and T.~Huang, ``Wide
  activation for efficient and accurate image super-resolution,'' {\em arXiv
  preprint arXiv:1808.08718}, 2018.

\bibitem{srgan}
C.~Ledig, L.~Theis, F.~Husz{\'a}r, J.~Caballero, A.~Cunningham, A.~Acosta,
  A.~Aitken, A.~Tejani, J.~Totz, Z.~Wang, {\em et~al.}, ``Photo-realistic
  single image super-resolution using a generative adversarial network,'' in
  {\em Proceedings of the IEEE conference on computer vision and pattern
  recognition}, pp.~4681--4690, 2017.

\bibitem{esrgan}
X.~Wang, K.~Yu, S.~Wu, J.~Gu, Y.~Liu, C.~Dong, Y.~Qiao, and C.~Change~Loy,
  ``Esrgan: Enhanced super-resolution generative adversarial networks,'' in
  {\em Proceedings of the European Conference on Computer Vision (ECCV)},
  pp.~0--0, 2018.

\bibitem{ma2020structure}
C.~Ma, Y.~Rao, Y.~Cheng, C.~Chen, J.~Lu, and J.~Zhou, ``Structure-preserving
  super resolution with gradient guidance,'' in {\em Proceedings of the
  IEEE/CVF Conference on Computer Vision and Pattern Recognition},
  pp.~7769--7778, 2020.

\bibitem{ranksrgan}
Z.~Wenlong, L.~Yihao, C.~Dong, and Y.~Qiao, ``Ranksrgan: Generative adversarial
  networks with ranker for image super-resolution,'' {\em IEEE Transactions on
  Pattern Analysis and Machine Intelligence}, 2021.

\bibitem{hcflow}
J.~Liang, A.~Lugmayr, K.~Zhang, M.~Danelljan, L.~Van~Gool, and R.~Timofte,
  ``Hierarchical conditional flow: A unified framework for image
  super-resolution and image rescaling,'' in {\em Proceedings of the IEEE/CVF
  International Conference on Computer Vision (ICCV)}, pp.~4076--4085, October
  2021.

\bibitem{gan}
I.~Goodfellow, J.~Pouget-Abadie, M.~Mirza, B.~Xu, D.~Warde-Farley, S.~Ozair,
  A.~Courville, and Y.~Bengio, ``Generative adversarial nets,'' in {\em
  Advances in neural information processing systems}, pp.~2672--2680, 2014.

\bibitem{wls}
Z.~G. Li, J.~H. Zheng, and S.~Rahardja, ``Detail-enhanced exposure fusion,''
  {\em IEEE Transactions on Image Processing}, vol.~21, no.~11, pp.~4672--4676,
  2012.

\bibitem{fwls}
D.~Min, S.~Choi, J.~Lu, B.~Ham, K.~Sohn, and M.~N. Do, ``Fast global image
  smoothing based on weighted least squares,'' {\em IEEE Transactions on Image
  Processing}, vol.~23, no.~12, pp.~5638--5653, 2014.

\bibitem{wang2019detail}
Q.~Wang, W.~Chen, X.~Wu, and Z.~Li, ``Detail-enhanced multi-scale exposure
  fusion in yuv color space,'' {\em IEEE Transactions on Circuits and Systems
  for Video Technology}, vol.~30, no.~8, pp.~2418--2429, 2019.

\bibitem{kou2017intelligent}
F.~Kou, Z.~Wei, W.~Chen, X.~Wu, C.~Wen, and Z.~Li, ``Intelligent detail
  enhancement for exposure fusion,'' {\em IEEE Transactions on Multimedia},
  vol.~20, no.~2, pp.~484--495, 2017.

\bibitem{wei2016local}
Z.~Wei, C.~Wen, and Z.~Li, ``Local inverse tone mapping for scalable high
  dynamic range image coding,'' {\em IEEE Transactions on Circuits and Systems
  for Video technology}, vol.~28, no.~2, pp.~550--555, 2016.

\bibitem{farbman2008edge}
Z.~Farbman, R.~Fattal, D.~Lischinski, and R.~Szeliski, ``Edge-preserving
  decompositions for multi-scale tone and detail manipulation,'' {\em ACM
  Transactions on Graphics (TOG)}, vol.~27, no.~3, pp.~1--10, 2008.

\bibitem{gif}
K.~He, J.~Sun, and X.~Tang, ``Guided image filtering,'' in {\em European
  conference on computer vision}, pp.~1--14, Springer, 2010.

\bibitem{wgif}
Z.~Li, J.~Zheng, Z.~Zhu, W.~Yao, and S.~Wu, ``Weighted guided image
  filtering,'' {\em IEEE Transactions on Image processing}, vol.~24, no.~1,
  pp.~120--129, Jan. 2015.

\bibitem{fsrcnn}
C.~Dong, C.~C. Loy, and X.~Tang, ``Accelerating the super-resolution
  convolutional neural network,'' in {\em European conference on computer
  vision}, pp.~391--407, Springer, 2016.

\bibitem{espcn}
W.~Shi, J.~Caballero, F.~Husz{\'a}r, J.~Totz, A.~P. Aitken, R.~Bishop,
  D.~Rueckert, and Z.~Wang, ``Real-time single image and video super-resolution
  using an efficient sub-pixel convolutional neural network,'' in {\em
  Proceedings of the IEEE conference on computer vision and pattern
  recognition}, pp.~1874--1883, 2016.

\bibitem{drrn}
Y.~Tai, J.~Yang, and X.~Liu, ``Image super-resolution via deep recursive
  residual network,'' in {\em Proceedings of the IEEE conference on computer
  vision and pattern recognition}, pp.~3147--3155, 2017.

\bibitem{loss}
H.~Zhao, O.~Gallo, I.~Frosio, and J.~Kautz, ``Loss functions for image
  restoration with neural networks,'' {\em IEEE Transactions on Computational
  Imaging}, vol.~3, no.~1, pp.~47--57, 2016.

\bibitem{enhancenet}
M.~S. Sajjadi, B.~Scholkopf, and M.~Hirsch, ``Enhancenet: Single image
  super-resolution through automated texture synthesis,'' in {\em Proceedings
  of the IEEE International Conference on Computer Vision}, pp.~4491--4500,
  2017.

\bibitem{singan}
T.~R. Shaham, T.~Dekel, and T.~Michaeli, ``Singan: Learning a generative model
  from a single natural image,'' in {\em Proceedings of the IEEE International
  Conference on Computer Vision}, pp.~4570--4580, 2019.

\bibitem{Fuoli}
D.~Fuoli, L.~Van~Gool, and R.~Timofte, ``Fourier space losses for efficient
  perceptual image super-resolution,'' in {\em Proceedings of the IEEE/CVF
  International Conference on Computer Vision (ICCV)}, pp.~2360--2369, October
  2021.

\bibitem{srflow}
A.~Lugmayr, M.~Danelljan, L.~Van~Gool, and R.~Timofte, ``Srflow: Learning the
  super-resolution space with normalizing flow,'' in {\em European Conference
  on Computer Vision}, pp.~715--732, Springer, 2020.

\bibitem{yang2017deep}
W.~Yang, J.~Feng, J.~Yang, F.~Zhao, J.~Liu, Z.~Guo, and S.~Yan, ``Deep edge
  guided recurrent residual learning for image super-resolution,'' {\em IEEE
  Transactions on Image Processing}, vol.~26, no.~12, pp.~5895--5907, 2017.

\bibitem{sftgan}
X.~Wang, K.~Yu, C.~Dong, and C.~Change~Loy, ``Recovering realistic texture in
  image super-resolution by deep spatial feature transform,'' in {\em
  Proceedings of the IEEE Conference on Computer Vision and Pattern
  Recognition}, pp.~606--615, 2018.

\bibitem{li2020learning}
M.~Li, Z.~Zhang, J.~Yu, and C.~W. Chen, ``Learning face image super-resolution
  through facial semantic attribute transformation and self-attentive structure
  enhancement,'' {\em IEEE Transactions on Multimedia}, vol.~23, pp.~468--483,
  2020.

\bibitem{perceptual}
J.~Johnson, A.~Alahi, and L.~Fei-Fei, ``Perceptual losses for real-time style
  transfer and super-resolution,'' in {\em European conference on computer
  vision}, pp.~694--711, Springer, 2016.

\bibitem{ntire}
R.~Timofte, E.~Agustsson, L.~Van~Gool, M.-H. Yang, and L.~Zhang, ``Ntire 2017
  challenge on single image super-resolution: Methods and results,'' in {\em
  Proceedings of the IEEE Conference on Computer Vision and Pattern Recognition
  Workshops}, pp.~114--125, 2017.

\bibitem{adam}
D.~P. Kingma and J.~Ba, ``Adam: A method for stochastic optimization,'' {\em
  arXiv preprint arXiv:1412.6980}, 2014.

\bibitem{lugmayr2020ntire}
A.~Lugmayr, M.~Danelljan, and R.~Timofte, ``Ntire 2020 challenge on real-world
  image super-resolution: Methods and results,'' in {\em Proceedings of the
  IEEE/CVF Conference on Computer Vision and Pattern Recognition Workshops},
  pp.~494--495, 2020.

\bibitem{ji2020real}
X.~Ji, Y.~Cao, Y.~Tai, C.~Wang, J.~Li, and F.~Huang, ``Real-world
  super-resolution via kernel estimation and noise injection,'' in {\em
  Proceedings of the IEEE/CVF Conference on Computer Vision and Pattern
  Recognition Workshops}, pp.~466--467, 2020.

\bibitem{li2003adaptive}
Z.~Li, ``Adaptive basic unit layer rate control for jvt,'' in {\em JVT 7th
  Meeting, Pattaya, Mar2003}, 2003.

\bibitem{liu2006novel}
Y.~Liu, Z.~G. Li, and Y.~C. Soh, ``A novel rate control scheme for low delay
  video communication of h. 264/avc standard,'' {\em IEEE Transactions on
  Circuits and Systems for Video Technology}, vol.~17, no.~1, pp.~68--78, 2006.

\end{thebibliography}

\end{document}